# Astrometry 1960-80:
# from Hamburg to Hipparcos

By Erik Høg, 2014.08.06
Niels Bohr Institute, Copenhagen University

ABSTRACT: Astrometry, the most ancient branch of astronomy, was facing extinction during much of the 20$^{th}$ century in the competition with astrophysics. The revival of astrometry came with the European astrometry satellite Hipparcos, approved by ESA in 1980 and launched 1989. Photon-counting astrometry was the basic measuring technique in Hipparcos, a technique invented by the author in 1960 in Hamburg. The technique was implemented on the Repsold meridian circle for the Hamburg expedition to Perth in Western Australia where it worked well during 1967-72. This success paved the way for space astrometry, pioneered in France and implemented on Hipparcos. This report gives a detailed personal account of my life and work in Hamburg Bergedorf where I lived with my family half a century ago.

The report has been published in the proceedings of the meeting in Hamburg on 24 Sep. 2012 of Arbeitskreis Astronomiegeschichte of the Astronomische Gesellschaft.

Gudrun Wolfschmidt is editor of the book:
   *"Kometen, Sterne, Galaxien – Astronomie in Hamburg"*
zum 100jährigen Jubiläum der Hamburger Sternwarte in Bergedorf. Nuncius Hamburgensis, Beiträge zur Geschichte der Naturwissenschaften, Band 24, Hamburg: tredition 2014

The introduction and the table of contents of the book as in early 2013:
`https://dl.dropbox.com/u/49240691/100-HS3intro.pdf`

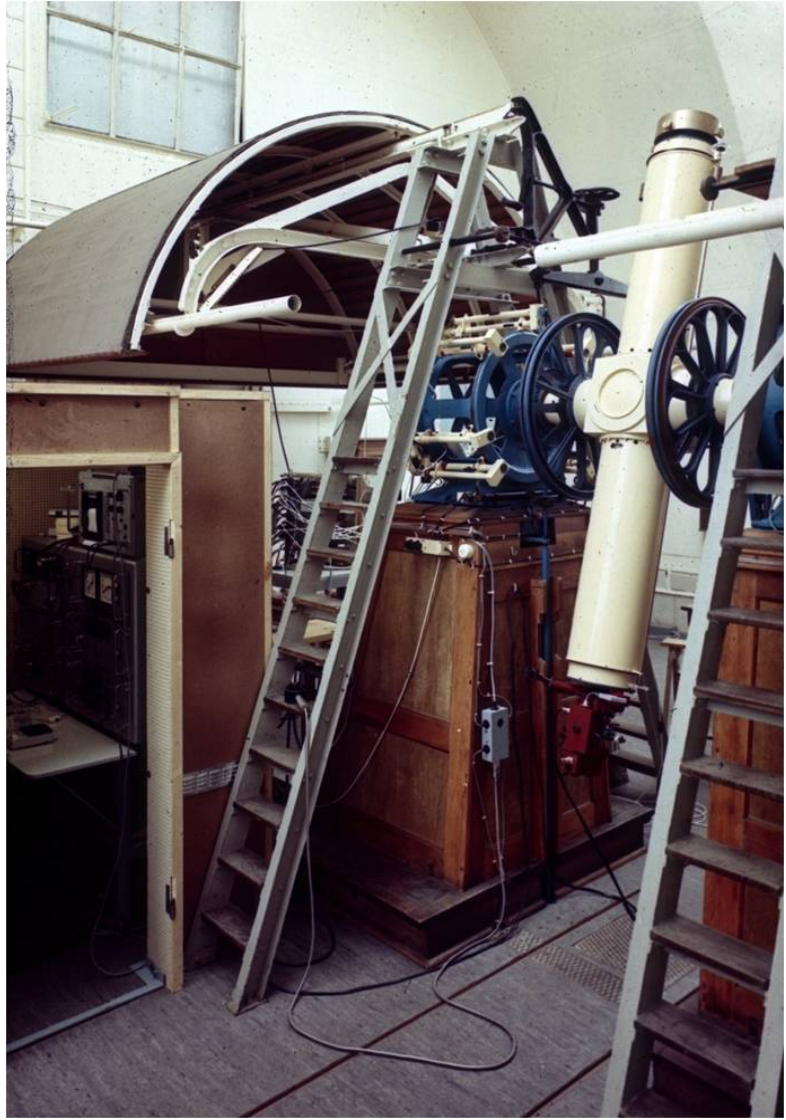

**Figure 12**

**Hamburg 1966 - The Repsold meridian circle ready for Perth**

The observer set the telescope to the declination as ordered by the assistant sitting with the star lists in the hut at left. He started the recording when he saw the star at the proper place in the field of view.

Photo: Wilhelm Dieckvoss

# Astrometry 1960-80: from Hamburg to Hipparcos

By Erik Høg,
Niels Bohr Institute, Copenhagen University
**email:** ehoeg@hotmail.dk

ABSTRACT: Astrometry, the most ancient branch of astronomy, was facing extinction during much of the 20$^{th}$ century in the competition with astrophysics. The revival of astrometry came with the European astrometry satellite Hipparcos, approved by ESA in 1980 and launched 1989. Photon-counting astrometry was the basic measuring technique in Hipparcos, a technique invented by the author in 1960 in Hamburg. The technique was implemented on the Repsold meridian circle for the Hamburg expedition to Perth in Western Australia where it worked well during 1967-72. This success paved the way for space astrometry, pioneered in France and implemented on Hipparcos. This report gives a detailed personal account of my life and work in Hamburg Bergedorf where I lived with my family half a century ago.

## Contents

Here follows a short overview of the development leading to the Hipparcos and Gaia satellites while the appendix gives a detailed personal account of my life and work in Hamburg half a century ago.

My presentation at the meeting in Hamburg (Høg 2013) was similar to the one I gave at the conference in Tartu University in 2011 on the occasion of the 200$^{th}$ anniversary of the Tartu Observatory, and the proceedings from that conference contains my talk (Høg 2011a).

**Photon-counting astrometry in Hamburg led to Hipparcos and Gaia**

On 1 October 1958 I arrived in Bergedorf with funding for 10 months from the *Deutsche Akademische Austauschdienst*. I wanted to get into astrophysics but my life developed very differently. I was pulled away from astrophysics when I invented a new technique for astrometry on 22 July 1960. The 10 months became 15 years, 1958-73.

The new proposed astrometric technique, photon-counting astrometry (Høg 1960) found immediate recognition, but it took seven years to implement it for the meridian circle in Bergedorf. The technique was digital and applied slits and photon counting. The photon counts were recorded on punched tapes for later reduction in the digital computer GIER. This was the first application of photon counting in astrometry and the counts were tagged with high-precision timing in Universal Time. The computer GIER of Danish manufacture was installed in 1964. GIER was the first computer on site in Bergedorf and after 1967 the first in the Perth Observatory in Western Australia. The expedition to Perth 1967-72 with eight observers from Hamburg and four from Perth was a great success resulting in the *Perth70 Catalogue* in 1976 with accurate positions and photometry for 24900 stars.

The technique with slits and photon counting proposed in 1960 was quickly adopted in France where the V-shaped slit system was called *"une grille de Høg"* and the technique was applied in the French work on space astrometry. The idea about astrometry from space was a French project beginning in 1964 - a great French vision it truly was, especially due to Professor Pierre Lacroute (1967, 1974).

In those years, I myself and many others considered the French design of an astrometric satellite to be completely unrealistic, but my opinion about astrometry from space changed in 1975. I was invited by ESA to join a small team for a mission study of space astrometry. At the first meeting in Paris on 14 October we were encouraged to forget about the specific French design and only to think about how we could make best use of space technology for our science, astrometry. I returned to Denmark and six weeks later I had sent three reports about a quite new design of the space instrument and mission. This quick development was founded on the new ideas from Copenhagen in the 1950s about astrometry, electronic instrumentation and digital computing and after these ideas had grown on the active and fertile grounds for astrometry in Bergedorf and Copenhagen/Brorfelde in the 1960s.

The subsequent technical and scientific studies based on the new design resulted in the approval of the Hipparcos mission in 1980 (see Høg 2011a and b) which observed 1989-93. The astrometric and photometric results from Hipparcos were published in 1997. They were improved for the bright stars in 2007 and provided a new basis for astrophysics. For example, 719 stars obtained distances with a standard error of 1% or better while only one star, the Sun, was known with a better accuracy before Hipparcos.

In 1992, during the Hipparcos mission, I proposed the design principles for a new astrometric mission Roemer (Høg 1993) which would be nearly a million times more powerful than Hipparcos. Such a mission was approved as cornerstone mission of ESA in 2000 and is due for launch in October 2013. This mission named Gaia is expected to obtain distances with 1% accuracy or better for 11 million stars. It has been proposed to give it a new name RGaia after launch, just as MAP was renamed WMAP. This would recognize the role of Roemer in actually starting the work towards Gaia and even in the right direction of direct scanning with CCDs, i.e. without the interferometry which was however very essential in the original GAIA proposal of 1993, see Høg (2011c).

Another fruit of my experiences from Hamburg and Perth was the development of the automatic Carlsberg Meridian Circle, placed on La Palma since 1984 and still in operation, see more in the appendix. It has been under remote control since 1997, one of the first robotic astronomical telescopes.

After leaving the Gaia Science Team in 2007, I have written a dozen reports about the history of astrometry and optics during the past 2000 years (Høg 2008a and 2011d) and I have given talks on the subject at two dozen places. After one of the talks a comment was this: *"You said you wanted to get into astrophysics fifty years ago, but by returning to revolutionize astrometry you have done more for astrophysics than you could have done as an astrophysicist"*.

**Acknowledgements**: I am grateful to William van Altena, Craig Bowers, Andreas Burkert, Markiyan S. Chubey, Dafydd Evans, Claus Fabricius, Gerry Gilmore, Christian Gram, Rustem I. Gumerov, Klaus von der Heide, Ilse Holst, Birgitte Høg, Jean Kovalevsky, Bernd Loibl, Leslie Morrison, José Luis Muiños, Peter Naur, Finn Verner Nielsen, Jørgen Otzen Petersen, Werner Pfau, Gennady I. Pinigin, Jochen Schramm, P. Kenneth Seidelmann, Chris Sterken, Anke Vollersen, Roderick Willstrop, Uwe Wolter and Norbert Zacharias for information and comments to previous versions of this report. When in 2009 I first met Rajesh Kochhar, now president of the IAU commission for the history of astronomy, he insisted that I should write a book, my scientific biography. Without the encouragement from him, Gertie Skaarup, Gudrun Woldschmidt and last, but not least from my dear wife Aase I could not have brought all this together.

**APPENDIX:**
# Life and work in Hamburg

**Preample**
This report contains mainly biographical notes about my work and life in Bergedorf from 1958 to 73. It shall give an impression of the life at the Hamburger Sternwarte in this period as seen from my perspective, and I refer to the book by Schramm (2010) for a general historical account of this period in Bergedorf. The scientific work around the expedition to Perth including the instrument development and the results has been described in numerous papers of the time as shall be mentioned and references be given. I will here focus on the daily life, on how I was welcomed 13 years after the end of the Nazi occupation of my country, on how the scientific ideas developed, and on some of my private life in Germany. The focus is on my positive experiences, but for a better understanding of the scientific process I include some of the darker sides of a life among astronomers.

# 1. Towards astrophysics

I left my home country, eager to learn and to work with astrophysics in Hamburg, and at the age of 26 years I was of course on the watch for girls, especially the dream girl I could spend the rest of my life with.

Leaving Denmark before midnight was well planned when I travelled to Germany by the night train on 30 September 1958. When I did so I should not pay tax for the last quarter of the year and I had always learned to think economically. The same did the Danish authorities who tried to collect the tax, but in vain. I arrived early in the day at the Hamburger Sternwarte in Bergedorf where I worked for the next 15 years, but that was certainly not planned. I came for a stay of just ten months funded by the *Deutsche Akademische Austauschdienst.*

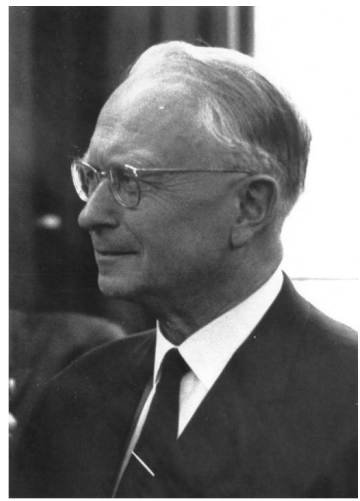

**Figure 1** Otto Heckmann. – Photo by WD
Key to sources of figures: CB: Craig Bowers, WD: Wilhelm Dieckvoss, IH: Ilse Holst, BL: Bernd Loibl, EH: Erik Høg, BS: Bengt Strömgren, and HV: Hans-Heinrich Voigt.

The director of the Bergedorfer Sternwarte, Professor Otto Heckmann (Figure 1), had visited Copenhagen a few years before in his effort to establish again the contact with Scandinavian astronomers after the World War. There were meetings of a few days in those years, called Baltic meetings, with lectures and discussions, in Copenhagen, Lund, Hamburg etc., and I saw the grand old men of astronomy, Bertil Lindblad, Schalen, Holmberg, Larink, Dieckvoss and Unsöld – they were not even old then. In the Jahresberichte for 1957, e.g., a working meeting in September in Lund is mentioned where Swedish, Danish and German astronomers participated, with ten from Hamburg.

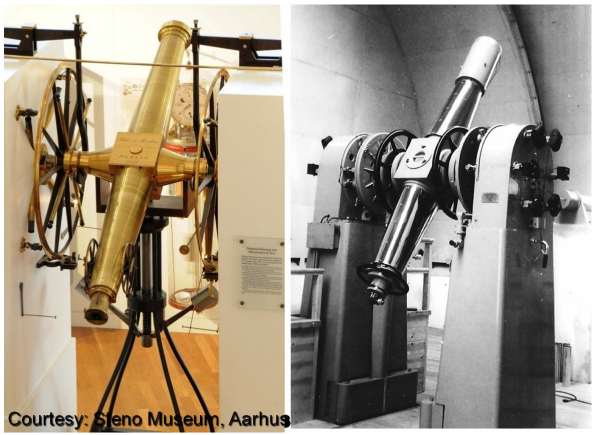

**Figure 2**   Left: the Copenhagen meridian circle as used in 1925 – by Pistor & Martins, Berlin, 1859. Right: the new meridian circle by Grubb Parsons, Newcastle, installed in 1953 at Brorfelde, 50 km to the west of Copenhagen. – EH

One of my young colleagues in Bergedorf told me later that Heckmann had reported of his first visit to Copenhagen, which could have been 1955. He had met a young guy working with the meridian circle in Brorfelde (Figure 2) and *"he seemed not to know really what he was doing"*, my colleague told. That must have been me, because Peter Naur, my astronomy teacher (see the Figure 3 and the later section), had told me to show Heckmann a series of photographic observations of the Polarissima. I had taken them with the new meridian circle at Brorfelde in order to monitor the stability of the instrument. There were only some fifteen measures in one coordinate and I was pondering whether one could say that there was a drift or not. I believe Heckmann advised me to get some more measures, and that was the first impression Heckmann had of me.

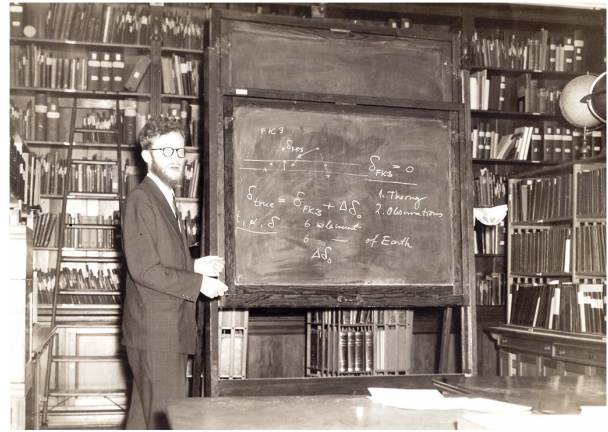

**Figure 3**   Peter Naur in the Bergedorf library in January 1958 giving a talk on an astrometric subject, systematic errors in the declination system. - WD

On that morning of 1st October I was welcomed by Heckmann in his office. I was shown around in the observatory, saw the big refractor, 60 cm aperture and the big Schmidt, the biggest telescopes I had ever seen. I asked Heckmann whether the micrometer for the refractor was available for measurement of double stars. He immediately called Herrn[1] Schultz, head of the mechanical workshop and asked him to make the micrometer ready. I measured some double stars proposed by Einar Hertzsprung and they were published in *Journal des Observateur*.

I was quite impressed and grateful that the director took such immediate action for a youngster like me and mentioned it in a letter to Peter Naur. He thought it was very natural and that Heckmann should be happy to see someone using the instruments.

Surely things changed, especially towards the end of my fifteen years in Bergedorf. Professor Alfred Weigert then received me sitting in the director chair with his legs stretched and feet on the desk, though not quite towards my face. He was accordingly frosty about my new plans for a revival of meridian astronomy in Bergedorf with the meridian circle to be returned from Australia to Bergedorf. These plans were for me the most natural consequence of the success in Australia, they were within the rich tradition in Bergedorf for meridian astronomy – I felt it was my duty to make such plans and I liked to do so and discussed them, especially with Gerhard Holst (1934-2000) whom I saw as a central person in the project. I submitted the plans to the institute council (Institutsrat) in May 1972 (on 12+7 pages) where they were treated with Gerhard Holst, me and many others as listeners since the meetings were open. The decision was a firm rejection.

---

[1] In Germany, a surname was often preceeded with "Dr.", "Herr", "Frau" or "Fräulein". I am not able to define precisely when it was done and when not, but I will write what I feel to be natural here.

I should also mention that I had pleasant talks with Weigert at other informal occasions: about Denmark, about my ideas of a new type of meridian circle, the glass meridian circle, and about our timing of the optical Crab pulsar with the big refractor. I wrote a popular article about the Crab pulsar, Høg & Lohsen (1971) encouraged by Weigert. A letter of 24 February 1971 from the director, Weigert, to von der Heide closes with the encouraging words: *"Auf gar keinen Fall sollten Sie den Mut sinken lassen; ich bin überzeugt, dass Sie die Expedition zu einem befriedigenden Abschluss bringen werden."*

In retrospect, I am happy that Weigert was negative. The rejection of my plans led to their realization in Denmark and it meant that I left no obligations or opportunities behind me in Hamburg. I could return to Denmark in September 1973 where my ideas and expertise were wanted and where the very good mechanical and electronic workshops were crucial for the further development of astrometry on the ground and in space.

**Original Schmidt plate in my table**

In October 1958 I was installed in the Lippert building where I had my office during all the years. My first table in a big room was facing a window towards west. In the drawer I found a big glass plate, about 50 cm diameter and one centimeter thick. It was the original Schmidt plate manufactured by Bernhard Schmidt twenty years before, as I could read from a diagram with a curve plotted, also in the drawer. The plate and diagram are now exposed in the Schmidt exhibition. One evening when I was working there as I often did, Heckmann came to see me and his interest pleased me greatly.

The other inhabitants of the Lippert were Dr. J.W. Tripp who worked for Heckmann and Professor A. Wachmann who's office was close to the entrance. I can still hear the gentle slam in my ear of the entrance door behind when I was some distance off, I used a very precise push in passing the heavy door. But the slam was not quite to the liking of Wachmann, it was of course a bit louder than when he entered and closed the door "normally".

After a year when Dr. Tripp had left, I moved into Schwassmann's office towards the east where I stayed for the rest of the 15 years. I sometimes saw a deer or rabbit and once even a blonde beauty on the grass. She came smiling to the window when I opened, Ilse was her name and she was picking flowers with Gerhard Holst whom she had just married. They soon became my collaborators on the meridian circle project and we have stayed in contact ever since. Several years I shared the nice office with F. von Fischer-Treuenfeld who worked with me, especially on the determination of the division corrections of the meridian circle thus earning his doctor's degree, see von Fischer-Treuenfeld (1968).

The telescopes were spread over the area of the observatory in Bergedorf, each one in a building with several offices. There were no telescopes in the large main building which housed the library and many offices. In three large houses close to the main building the astronomers and employees lived with their families. This distributed structure limited the daily communication between astronomers, but we met to the weekly seminar in the library on Saturday morning.

In 1958 I soon found my way to the Schmidt telescope building where some of the young people used to meet for coffee once a week, or was it every afternoon? The names changed with the years, the first were Kristen Rohlfs, Brosterhuis, Oswald, Schücking, Herczeg and Tripp. We met there in the office reserved for Walter Baade. It was not used otherwise because Baade died in Göttingen in 1960 before he could return to the Hamburg observatory from where he had left for America in 1931. We somewhat hesitantly invaded the big beautiful office with good chairs without permission. At first it had a fresh smell of wood from the Schmidt plate archive in the office, but that smell soon faded behind the cigarette smoke although we feared Heckmann's reprimand. He was a non-smoker and nobody smoked in his presence, but he was busy with ESO and he luckily never appeared in the smoking stinking Schmidt or ever said anything.

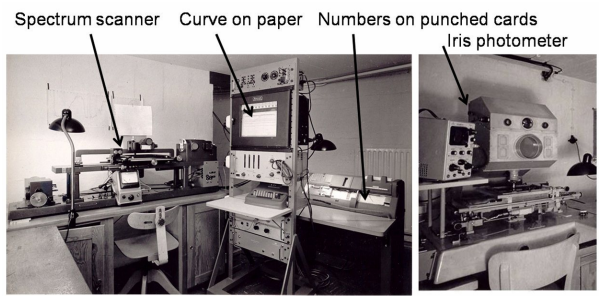

**Figure 4**   Digitization in 1960 in Bergedorf, located in the Sonnenbau, lower floor. - Left: Spectra of stars were photographed and then measured with a scanner. The measurements came hitherto as a curve on paper, but now also as numbers on punched cards. Right: The diameter of the round stellar image was measured with the "iris-photometer" and the number was punched. – WD

**Digitize plate measurement**

The Schmidt telescope was either used to take direct plates for a photometric survey of open clusters or plates with low-dispersion spectra. The clusters were measured with an iris photometer, called the Haffner photometer after the designer. The reading of the iris was noted by hand on the star list. When I saw that, I soon convinced them that the instrument could be easily digitized. A potentiometer was attached in 1960, the voltage was converted by an analog-to-digital converter and the numbers punched on cards (Figure 4).

All the components were commercially available. Some of us went to the big Hannover Messe a day every year. That was my source of information for several years about detectors and digital equipment. We convinced the director, Professor Hans Haffner, that he should support these visits. As Heckman became director of ESO in 1962, Haffner had taken over and he was a good support for astrometry. I had the great pleasure with my wife to visit Mrs Haffner in 2005 in Würzburg where she lived a good senior life.

I soldered the components myself having the experience from the time in Brorfelde. So I did during the following years with much of the equipment for the meridian circle. I built e.g. power supplies from basic components since they were not commercially available, or too expensive for my low budget. My electronics workshop was established in 1962 on the upper floor of the building "Sonnenbau", at the middle of the observatory area. There was no electronic engineer yet, not before Dipl. Phys. F. von Fischer-Treuenfeld was imployed from 1 October 1964 and he stayed until 31 December 1967. This imployment of an engineer came 2.5 years later than planned because we had waited for another person, who finally chose another job. Herr Ingenieur (grad.) Fritz Harde was later hired and worked with me for two years (1 August 1966-31. July 1968).

Spectra could be scanned on the spectro-photometer and I envisaged using the low-dispersion Schmidt spectra for automatic classification. I saw that as my way into astrophysics, but never got very far with the work because I got the idea about photoelectric astrometry. And perhaps it was a bit too early for automatic analysis of spectra considering the computer we had in 1959, an *IBM 650* located in the university 20 km drive from Bergedorf.

With that computer, I developed an idea I had got in Brorfelde, a new mathematical method for the determination of corrections to the lines of a divided circle on a meridian circle and I submitted my first scientific publication to *Astronomische Nachrichten* in April 1960 (Høg 1961b). The method required a simpler set up of the microscopes for measurements and it made use of least-squares solutions. It required much more computing than previous methods. But the computing was no more a problem, the results were more accurate, and the method was in fact soon applied at all active meridian circles (see Rapaport 1985).

A few years later Tibor Herczeg came to me with a young student, Michael Grewing, because they thought of using the digitized spectrometer for a project, but I did not recommend that. Many years later, in 1981 when Grewing had become director of the Tübingen institute of astronomy he was one of the first to support the Tycho project on Hipparcos and from about 1984 a group in Tübingen was one of the corner stones in the Tycho data analysis. Andreas Wicenec from the group in Tübingen where he began as a young student was vital for the production of the *Tycho-2 Catalogue* completed in 2000. He later received all the original Tycho observations and intended to include them in the ESO archives in Garching.

The digitized spectrometer in Bergedorf was finally used by a student, Herr Leichner, from Professor Th. Schmidt-Kaler's institute in Bochum. I was later told that Leichner did not like the experimental work, but he later turned out to be a good theoretical scientist.

## 2. No return to Denmark

A tenure position was announced in 1959 as leader of a new radiation laboratory in the Danish city Odense, similar to the laboratory at the Finsen Institute in Copenhagen where I worked during my soldier time. I talked with my former boss there, Ambrosen, about this possibility to come back to Denmark. He did not think that such a position was something for me and wisely advised me to pursue a career as astronomer.

A tenure position was then announced at the astronomical institute of Aarhus in Denmark and I went to see the professor, Mogens Rudkjøbing (1915-2000), whom I knew from his course in astrophysics in Copenhagen. We looked at the spectrum scanner and agreed we could consider digitising it. Back in Hamburg I designed an equipment for the purpose and inquired at a Danish firm for the price, in Danish money because it should be delivered to Aarhus in case of a decision to purchase. I received some time later a letter from the Professor telling that the firm had contacted him, and he blamed me for having arranged with a firm *"behind his back"*. I was surprised at this reaction and mentioned it to Heckmann. He said with a smile that perhaps Aarhus was not the right place for me to go; two different characters. He was of course right as I could have known even before I applied in Aarhus, but I was eager to come back to Denmark. In Hamburg I could freely take initiatives, ask firms for offers etc. My direct way was welcome, especially with Heckmann.

A return to Copenhagen or Brorfelde was out of the question for many years, partly because of "two characters", the new professor and me, but also for another good reason. The new professor there since 1958, Anders Reiz (1915-2000), led the observatory into a fruitful future with astrophysics. But he was also eager to implement a photographic method of meridian observations in which much work had already been invested. In fact it became operational in the 1960s, in the hands of Svend Laustsen (*1927) who had been my fellow student, and it was quite successful.

## 3. The new astrometry

**Photon-counting astrometry in review**

Photoelectric astrometry with slits and photon counting may briefly be called "photon-counting astrometry". It was invented in 1960 as I will describe below. It was developed in Hamburg and I shall already here give an overview of the evolution and significance of this method up to the present time for fundamental *astrometry of large angles* on the ground and with a satellite.

Photoelectric astrometry was also developed in many other forms which are not subject of this presentation, e.g. for a pointing sensor or for slits, but using analog amplification of the photo current instead of counting, or for astrometry of small angles. The much more efficient CCD technique took over from photoelectric astrometry in the 1990s and will here be mentioned with a few examples only.

Svend Laustsen has told me recently that they had recognised my new photoelectric method as being the future, but it would have been out of the question to develop this method in Brorfelde at that time, and I fully understand.

The idea was however developed in the 1970s for the meridian circle in Besancon by M. Sauzeat. He presents *"an automatic micrometer using Høg's principle"* in Sauzeat (1974) and mentions it in two publications from 1971 and in his thesis from 1973. It was developed but not applied for routine observations. Such a slit micrometer was also developed for a transit instrument in Torino by Anderlucci et al. (1983) and was used for 1300 transit time observations of FK4 stars.

In the 1980s a slit micrometer was developed for a new meridian circle for Tokyo (Kühne, Miyamoto & Yoshizawa 1986). Observations of 33000 stars and of solar system objects were obtained during 1985-93 (Yoshizawa, Miyamoto et al. 1994).

In the USSR, photoelectric techniques using slit systems was pioneered at the Pulkovo Observatory by Pavlov (1946, 1956) using analog amplification of the photocurrent. Although photon-counting astrometry is the main subject of this section, some information shall be included here because of the pioneering character of the work and because an overview of astrometry in the USSR is hard to get. The observations were dedicated to the time service by transit instruments (Figure 5) and therefore only observation of transit times was needed, not of declinations, resulting in, e.g., a general catalogue of right ascensions for 687+120 stars by Pavlov et al. (1971), derived from 20 observed catalogues. This paragraph is based on recent correspondence with M.S. Chubey.

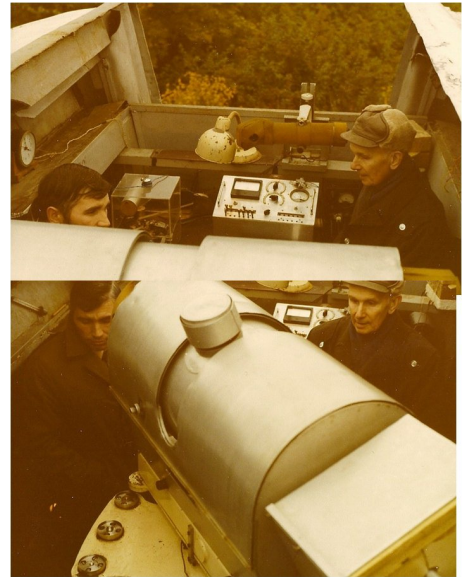

**Figure 5** Pulkovo Observatory in 1978. N.N. Pavlov at right pioneered photoelectric astrometry for time service with a transit instrument, G.I. Pinigin at left. The instrument is reversible as seen in the lower part of the figure. - EH

The Sukharev horizontal meridian circle (HMC) erected in Pulkovo Observatory in 1960 was equipped with a photoelectric micrometer using a fixed *"V-shape"* slits system. Observations of right ascensions of 188 stars and declinations of 224 stars for improvement of FK4 were made for two catalogs in 1970-1982: mean errors of 0.1 arcsec (Kirijan, Pinigin et al. 1984). In the 1980s the HMC was further developed and equipped with a new scanning slit micrometer with photon counting under computer control for observation of star positions, observing about 40 stars every clear hour. In 1987-1990 two catalogues of 911 star positions of FK4 and 170 reference stars in the 63 ERS fields were derived on the basis of 6000 observations. It was the first automatic telescope in USSR and produced four catalogues of good accuracy: mean errors of 0.08-0.10 arcsec and instrumental systematic errors on the level of FK5: 0.02-0.03 arcsec (Ayrov et al. 1984, Gumerov 1988, Gumerov et al. 1986).

In Nikolajev, the Axial Meridian Circle (AMC) was developed by G. Pinigin, O. Shornikov, and A. Shulga and put in operation in 1996. The AMC was equipped with a CCD camera in drift-scan mode, limit at 16th mag., under computer control with remote access, observing about 400 stars every clear night, Kovalchuk et al. (1997). Position observations of stars and Solar system objects with mean errors of 50 mas were obtained. Seven catalogues with a total of more than 350,000 star positions have been obtained up to 2008. These latter two paragraphs are based on recent correspondence with G.I. Pinigin and R.I. Gumerov.

**Danish, English and Spanish collaboration**

Prior to these activities, the first development of a slit micrometer with photon counting for mass production observations with a meridian circle after the one in Bergedorf began in Denmark in 1973. I returned from Hamburg in September and the design could begin for the meridian circle in Brorfelde, partly because the administrative authority had been transferred from the all-deciding professor to a democratic Institute Council.

Thanks to my experiences from Hamburg and Perth we were able to design and build a more advanced slit micrometer in the very good mechanical and electronics workshops in Brorfelde. That was the basis for a fully automatic meridian circle, a project led by L. Helmer (Helmer & Morrison 1985, Helmer, Evans et al. 2011) using the instrument from Brorfelde (Figure 2).

The instrument was moved to *Roque de los Muchachos* on La Palma in close collaboration with the *Royal Greenwich Observatory* and the *Real Instituto y Observatorio de la Armada* en San Fernando, and began operation in May 1984 under the name of the *Carlsberg Automatic Meridian Circle* (CAMC), observing about 500 stars every clear night. Annual catalogues were published, culminating in a combined catalogue of 181,000 star positions. The first meridian circle observations of Pluto were obtained and many faint minor planets were observed. A total of 25,848 positions and magnitudes of 184 Solar System objects may have been of more lasting value than the star positions, according to Morrison (2012, priv. comm.). The star catalogues were of such good quality at that time (mean errors of 0.06 seconds of arc) that they revealed systematic errors in the FK5. The *Carlsberg Meridian Catalogues La Palma* Numbers 1--11 (1999, on CD-ROM) were published by the three observatories.

The CAMC has been under remote control since 1997, one of the first robotic astronomical telescopes. It is still being operated, since 1999 with a CCD detector (Evans 2003, Evans et al. 2002) and 100,000 to 200,000 observations are obtained every night. Since 2006, the Spanish institute is the sole operator of the instrument.

A scanning slit micrometer with photon counting was built in Brorfelde and used on the *San Fernando Automatic Meridian Circle* (a twin instrument of the CAMC) on *El Leoncito* (Argentina) during the period Oct. 1997 to Sep. 1999 to produce a catalogue (HAMC) of 6192 positions and magnitudes of stars south of declination +40° and of 92 Solar System objects (Muiños, Mallamaci et al. 2001). Since Dec. 1999 the instrument has been operating with a CCD detector which was replaced by a larger one in 2009.

A CCD detector is based on semiconductors and is able to measure many stars simultaneously with very high quantum efficiency, ten times higher than the photo multiplier. The use of CCDs for drift scan on a meridian circle was planned by Stone & Monet (1989), perhaps the first such plan. Seven years later 2.2 million observations had been obtained, Stone et al. (1996).

It appears that the photoelectric method with slits and photon counting was used up to 1999, but it had become technically obsolete by 1990, thirty years after it was proposed in 1960. In summary, mass production astrometry on the ground with the photon-counting method was obtained with the five above mentioned meridian circles as follows: the Hamburg instrument in Perth 1967-87, the Danish instrument on La Palma 1984-99, the Japanese instrument in Tokyo 1985-93, the instrument in Pulkovo 1987-90, and the Spanish-Argentinian on El Leoncito 1997-99.

**Photon-counting astrometry in space**

In space, the photon-counting method was used on the first satellite dedicated to astrometry, the Hipparcos satellite which observed 1989-93. The results were published in the Hipparcos and Tycho Catalogues (Perryman, ESA 1997) containing 120,000 star with milliarcsecond astrometry and very accurate photometry and, in addition, one million Tycho stars with less accurate astrometry and two-colour photometry.

The Tycho raw data were reduced again resulting in the Tycho-2 Catalogue with 2.5 million stars, including proper motions derived from 100 years of astrometry (Høg et al. 2000). The Hipparcos raw data were also reduced a second time resulting in a new Hipparcos Catalogue with more accurate astrometry, especially for the bright stars (van Leeuwen 2007).

The second satellite dedicated to astrometry, Gaia, is due for launch by ESA in October 2013. It will be a million times more efficient than Hipparcos since it uses CCD detectors as proposed by the present author in 1992 for the Roemer satellite project (Høg 1993), a development described by Høg (2011c).

# 4. Return to astrometry with a great idea

**The new begin for photoelectric astrometry**

The year 1960 gave my work in Bergedorf a new turn. I had guests from Denmark, Svend Laustsen and others from Brorfelde. On 22 July we visited the photo-zenith telescope of Dr. Treusein in the *Stadtpark* of Hamburg. This is a special astronomical telescope which can measure a star just above us, i.e. in the zenith, and thereby the rotation of the earth is measured very precisely. A photographic plate is moved along with the star while the star image is exposed on the plate.

It was exactly in the discussion standing at that instrument that an idea struck me. My idea was to use a fixed plate with slits and let the star cross while recording the counts of a photo-multiplier tube placed behind the slits.

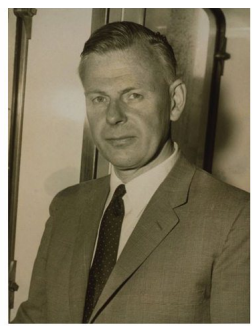 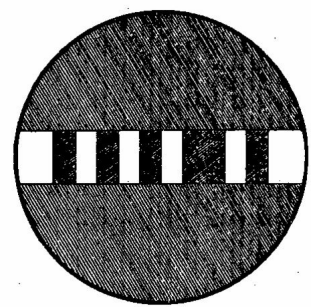

Bengt Strömgren (1957)

**Figure 6** Bengt Strömgren and the slit system he used in 1925 for experiments with photoelectric recording of transits. A photocell behind the slits gave a signal in which the transit time for each slit could be detected thus obtaining the right ascension of the star. - BS

I did not mention the idea on the spot, but I worked on it in the evening. The idea was based on an experiment at the Copenhagen meridian circle (Figure 2) with photoelectric astrometry in 1925 by Bengt Strömgren (1925, 1926, 1933), papers I had read as a student and kept in mind ever since. The experiment (Figure 6) began in January 1925, led by the 17 year old Bengt Strömgren and it was technically very advanced for its time - it is a mistake when I have in some reports of recent years called it "a modest experiment". On that background I had suddenly realized that photoelectric astrometry could best be implemented with photon counting (Figure 7).

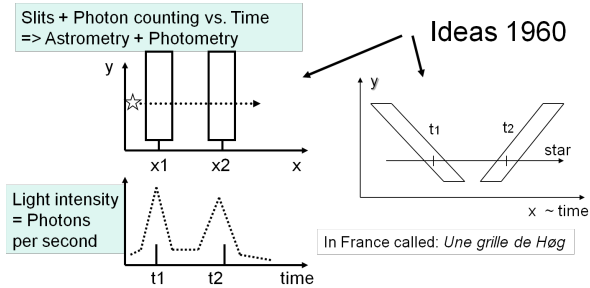

**Figure 7** The new ideas for photoelectric recording of transits. Left: a star crosses slits in the focal plane of the meridian circle and the photon counts are recorded with time. Right: with inclined slits also the vertical coordinate of the star is measured. Combined with circle reading, both right ascension and declination of the star will be measured. - EH

This was to be the first application of photon counting in astrometry and the counts were tagged with high-precision timing in Universal Time. Photon counting had already been used for stellar photometry in Cambridge since 1948 according to P. Naur and to R. Willstrop (2012, priv. comm.). The instrument was described and tested by Yates (1948) and results were published by Yates (1954).

I had learnt electronic counting techniques while I was a conscript soldier in 1956-57 working in a radiation laboratory. We collected rain and dust every day and measured the decay of the radioactive fall-out spread all over the northern hemisphere in those years from the atomic bomb tests in the atmosphere by the two super powers.

Many other things than photon counting and an accurate slit system had to be feasible before my idea would be realistic. Digital computing and programming were issues to which I will return. But in 1960 I saw the solution to the technical problem of storing the photon counts before they could be entered into a computer.

The photon counts in accurately timed intervals should be recorded at the telescope. Magnetic tape was unaffordable, but I had seen a tape punch (Figure 10) from Standard Electric Lorenz (SL614) at the Hannover Messe. It could punch 50 characters per second on 5-channel tape and it was not too expensive. That was the solution for our low budget, but years later when we started to use it at the meridian circle we realized that it was quite unreliable and it gave us much trouble. We changed to a Facit punch (PE1501), a Swedish product, which had seemed too expensive for us to begin with. It was also used at the computer, see Figure 15. It could do 150 characters per second on 8-channel tape and it was very reliable.

Our visit in the Stadtpark was on 22 July 1960 and my notes say that I worked up to 12 hours per day on that subject during the following time. I set out to study all aspects of the idea: optimal design, technical feasibility, theoretical basis and limitations e.g. due to the Earth atmosphere, expected accuracy and I published a number of papers during the following years.

For many years I noted every day in my calendar how many hours I had worked on my projects, division errors, spectrophotometry etc. and at the end of a year I made a statistics. I had complete freedom to do as I liked in Bergedorf, of course I always attended the seminar and enjoyed the social life especially in the big Schmidt. I once asked Heckmann if I could go on a trip with my parents and sister to the Weinstrasse, he smiled that I did not have to ask.

**First presentation of the idea in Jena**

The yearly meeting of the *Astronomische Gesellschaft* in the autumn of 1960 took place in Jena and Weimar in the then DDR. I presented the slit micrometer in a talk. One of the senior astronomers introduced me to a young man Werner Pfau who was also working on photoelectric astrometry. On the last day I walked with him through the *Goethe Museum* in Weimar looking at one and the other exhibit and most of the time talking about photoelectric astrometry. Many years later in 1998, I unexpectedly met Pfau again, now Professor Pfau who gave the opening speech to the conference in Gotha, celebrating the 200 year anniversary of the first European astronomy conference in 1798. I asked him if he remembered our walk through the *Goethe Museum*, and he answered: *"Of course I remember."* He showed me his small broken transit instrument from those times, exhibited at the meeting.

For the present review I asked Werner Pfau by email about his work in 1960. He answered that it was his student's thesis and he presented it at the meeting in Jena where we had met each other. He had implemented the method of Pavlov (1946, 56) for photoelectric astrometry with analog amplification and the accuracy was quite good, but only some 60 stars were observed. I will include more from his mail since it illustrates the working conditions in those years and also because there is quite some similarity with my conditions of work and, furthermore, we both wanted to get away from astrometry into astrophysics after our theses. Pfau succeed in this respect while I did not, but instead became associated with astrometry for the rest of my life thanks to fortunate circumstances.

Here is what Werner Pfau added, with his permission:

> *"Later on, I never understood how the staff members of the observatory could offer one of their students such a diploma thesis and how I could have accepted it. In the time before and after I always cared for the observatory clocks and, in itself, to electronically register these stellar transits was rather an interesting topic. But it laid absolutely sideline to the research field of the institute. There was not much astronomy in it and I had to do all the electronics work by myself. In these times, here in the GDR (DDR) you couldn't buy electronic equipment ready for use, but had to design and accomplish it yourself. I, therefore, turned to photographic and photoelectric photometry. This happened all the more because I soon became responsible for the institute's new outstation with the 90-cm telescope. Following the tradition of the institute, my interest turned to interstellar extinction and the spectra of the interstellar material."*

In 1990 Werner Pfau was appointed professor of astronomy and director of the university observatory in Jena and he was president (then Chairman) of the Astronomische Gesellschaft 1996-99.

Now back to 1960. The travel to Weimar was done in my old Volkswagen model 1953 accompanied by Elmar Brosterhuis, crossing the border at Helmstedt. On return we visited Naumburg, Wittenberg, and Magdeburg against strict orders from the DDR authorities. We saw the famous Naumburger Dom with the even more famous sculpture of the beautiful Ute from 1250. I still have a plate with Ute on the wall in our living room.

In Naumburg we found a place to stay over night which was also strictly forbidden. The lady of the house was very happy to talk with us from the west, and we talked and listened for many hours in the evening. The next morning we parted as old friends followed by her waving and her words: *"Auf Wiedersehen im vereinigten Deutschland"* (Au revoir in reunited Germany), which of course could not happen in our life times as we all knew – but it did happen! Germany was reunited in 1990, but of course we never saw the lady again, and she probably did not live to see what she had hoped for. But I thought of the lady and the first visit to Naumburg when I returned to show it all to my wife in 1998 after the conference in Gotha. Without any problems with the *Volkspolizei* in 1960, Brosterhuis and I returned to Hamburg.

**Publication and discussion**

I wrote a paper about the proposed photoelectric astrometry in the *Astronomische Abhandlungen der Hamburger Sternwarte*, Høg 1960. Heckmann helped me to improve my draft, sitting together an evening in his office. It is interesting to see that the paper has only seven citations in ADS in 2009, one of them is from France (Sauzeat 1974), although it created a lot of interest in Europe, especially in France, where they called my system of inclined slits *"une grille de Høg"*. An American astronomer, Professor

Larry Frederick from McCormick Observatory, told me a few years later that it inspired him to experiment with similar photoelectric astrometry with photon counting on a long-focus telescope, Frederick et al. (1975).

Clearly the proposal was considered to be important, but some colleagues did not agree that it would work well at all. One of them was the British astrometrist and astrophysicist R. d'E. Atkinson who insisted in a discussion I had with him in Bergedorf that the intensity scintillation of star light would kill the accuracy. He had seen these large intensity changes himself. This concern was justified in principle, but I could show theoretically that the scintillation would have only a very small effect, which he, however, did not believe.

My Dutch colleague G. van Herk was another declared sceptic in 1960, and wrote so in a paper to *Astronomical Journal*, where I countered him with a small letter (van Herk & van Woerkom 1961 and Høg 1961a). Van Herk soon became one of my good friends whom I often saw again, the last time in 1999 in his home in the Netherlands, shortly before he passed away. I often enjoyed showing him the enormous progress of astrometry with Hipparcos and Tycho, all in his lifetime and in spite of his innate pessimism, and he was happy to hear.

In 1961 I visited the astronomical institute in Tübingen twice because Professor H. Siedentopf, director of the institute had led investigations of scintillation and of stellar image motion. The image motion was the other atmospheric effect that would affect the astrometric accuracy with the proposed micrometer. The visits in Tübingen helped me to evaluate the effects of the atmospheric disturbances on the observations as later published by Høg (1961a, 1968, and 1970).

One of the astronomers from Tübingen was Hans Elsässer, then an expert in scintillation, who once visited me in my office in Hamburg on Heckmann's initiative. He became convinced that my theory in these matters was adequate.

My proposal was discussed at many occasions, by July 1962 (HS 1969) I had given talks in Weimar, Bonn, Heidelberg, Tübingen, Hamburg, Washington and Copenhagen about the instrument and the theory

I attended the IAU General Assembly in 1961 in Berkeley and especially the meetings about astrometry. I recall clearly that I thought I knew the solution to some of the problems in meridian astronomy, but I was too shy to ask for the word in a meeting. At the General Assembly in 1964 in Hamburg I was asked by the chairman of Commission 8 F.P Scott to give a short presentation which I did.

**Realize the idea for Australia**

It was soon clear that Heckmann wanted the idea implemented on the Repsold meridian circle. This instrument had been used to observe the AGK2A catalogue, about 20,000 reference stars for the photographic AGK2 catalogue for which the photographic plates were taken in Hamburg and Bonn about 1932. The then young observer from 1932, Dr. Johann von der Heide (1902-95), was now of mature age and was in the years around 1960 observing the AGK3R, the reference stars for the photographic plates of AGK3 which were being measured in Bergedorf.

Von der Heide was far from tired from all the meridian observations of thousands of stars, one by one. He was eager to lead an expedition with the meridian circle to Perth in Western Australia. The purpose was to observe the 20,000 Southern Reference Stars (SRS) which was an international undertaking decided in 1958 with 12 participating observatories in many countries.

The intention was to observe in Perth with the Hamburg meridian circle precisely as it was equipped for the ongoing AGK3R program, i.e. with visual observation of the star (Figure 8). The director of Perth Observatory, H.S. Spigl, was erecting a new observatory at a good site far outside Perth in view of the coming Hamburg expedition. According to Bowers (2012) Spigl visited Hamburg *"between 20th to 23rd July 1959 to see Dr. Heckmann, Dr. Deikvos and Dr. Larink"* and his notes describe the Hamburg

Observatory as *"This most efficient Observatory...."*.

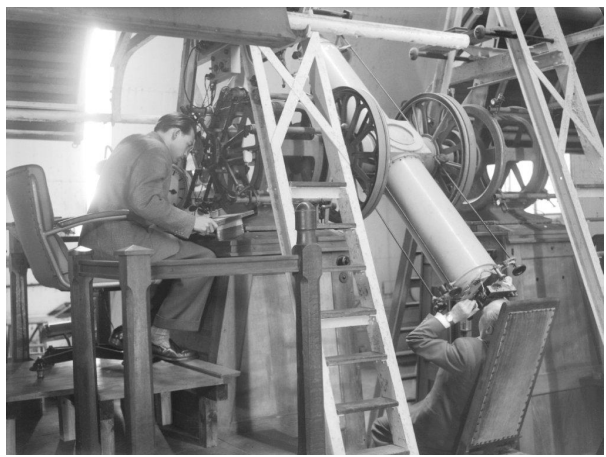

**Figure 8** The Repsold meridian circle about 1960. Gerhard Holst is reading the declination circle and Johann von der Heide is observing the star with the visual micrometer. - WD

Heckmann took Spigl and me on a sight seeing tour in the large Hamburg harbour, this happened a year before I had invented the photon-counting instrument. Spigl has probably also met von der Heide, since he was the coming leader of the expedition, but this did not happen in my presence. Von der Heide outlines his ideas for the observing program in Perth in a letter to W. Fricke of 14 February 1960.

The digitization for photometry and spectra had not given me many publications, but I won the respect and confidence of Heckmann that I had more to offer, and now I had come with the photon counting method for astrometry. An evening with clear sky, probably in late 1960, he asked me if I would join him on a trip to the Görde at Dannenberg to the east of Bergedorf where he was planning to build for the big Schmidt telescope at a site with darker sky; plans which never materialized.

He wanted to see the sky by night with own eyes, he said. Of course he also used the opportunity to talk with me about the expedition to Perth. Could he rely on me for many years to complete my plans for photoelectric astrometry? He knew that he was placing a great responsibility on me, but he did not mention that on the trip in the night, it only struck me much later. I remember clearly saying on that trip that I considered these ideas of photoelectric astrometry worth several years of my life.

On the way back to Bergedorf he asked if I would like to drive the car, a black Mercedes with steering-column gear lever which I had never tried before, neither had I ever driven such a big car. I liked to try and it went to the satisfaction of both of us.

I do not know how the decision to implement the photoelectric method for the expedition to Perth was taken, but I do not remember that I was directly involved in a decision process. It appears from the correspondence in Høg (2012), e.g. in a letter from von der Heide to Heckmann of 9 May 1961 that von der Heide was fully in favour of the automatic equipment.

In those years, however, I sometimes had the impression that it was not quite to von der Heide's liking. He was totally familiar with the traditional visual system, but knew only little about the electronics I was proposing. I did not feel his skepticism immediately and I believe he was quite impressed by me, also personally. It is natural that dissatisfaction could build up, especially as the expedition had to be postponed more than once because the implementation took more time than expected.

The instrument (Figures 10, 11, and 12) was shipped for Australia only in June 1967 when von der Heide was already about 65. The last postponement of the shipment from Hamburg by one month was very lucky, however. Had it not been for that, the instrument would have been stuck in Egypt for more than one year since the first planned ship was kept in the Great Bittersea due to the Seven-days-war begun on 5 June 1967.

**Heckmann launches after 102 days !**

Now back to 1960 and the implementation of the proposed photoelectric system. I defined the system for timing and recording on punched tape and Heckmann sent an application dated 1 November 1960 to the Deutsche Forschungsgemeinschaft (DFG). Thus only three months after I had the first idea has Heckmann taken his decision and launches the project. This is documented in HS 1969 and HS 1970 from which the following details are taken.

The application of two pages mentions at first Heckmann's application of 27 August 1959 for funding of an equipment for automatic spectrophotometry. He says that this application is now withdrawn because the equipment has already been built for other funds – as I have described above in the present paper and shown in Figure 4. Heckmann then comes to meridian circle astronomy and applies for 35000 DM for a *"photoelectric micrometer for a meridian circle"*. The application contains five annexes elaborated by me and including Høg (1960). Heckmann describes the properties and advantages of the new instrument.

He finally notes that the micrometer can be considered as the first step towards a further large development of a fully automatic meridian circle controlled by paper tape, and a block diagram of such an apparatus is shown in annex No. 5. The paper tape shall contain the right ascension, declination and magnitude of the stars in the sequence of observation.

Annex #5 about an automatic meridian circle closes literally as follows, translated from German:

> *"The cost of the electronics is estimated to 100,000 DM which is justified since the meridian circle itself is worth about 500,000 DM. If a meridian circle is placed in an astronomically good climate, e.g. in the southern hemisphere, such an automation would be especially profitable in order to increase the efficiency and save labour while the enormous observation material for about 100,000 observations per year is treated."*

Precisely this goal of 100,000 observations per year was adopted later for the automation carried out in Brorfelde and it was achieved by the Carlsberg Automatic Meridian Circle set up on la Palma in 1984, i.e. 24 years after Heckmann's application. The Perth expedition 1967-72 obtained 20,000 observations per year, unique for that time.

The application for the micrometer reports that offers to build the electronic equipment have been requested from four firms: Beckman Instruments, Kirem Gmbh, AEG and Siemens. The firm Kirem has already assured that they can build the equipment, but an offer will take another month.

# 5. Danish computer beats American by factor ten

The issue of computer is not discussed in Heckmann's application, but only mentioned once with the two words *"elektronische Rechenmaschine"*. With hindsight, this was very optimistic in 1960. Even three years later when the computer had to be chosen there was only one affordable computer on the market which could do what we needed. It was produced by a small Danish firm and it was ten times faster than its closest competitor produced by IBM. If we had been realistic in 1960 we would have seen that the project was unfeasible, but our ignorance, enthusiasm and wishful thinking kept us going! – These sentiments are often useful drivers in new science.

I followed the computer market closely in those years but my concerns were elsewhere. I did not at an early time see how critical the issue was with respect to computer hardware and software and to development of the computer programmes.

A plan found in a letter from von der Heide to Heckmann of 18 April 1961 (Høg 2012) foresees observations during two years beginning in late 1963. It does not mention a computer.

The data from the micrometer were punched on paper tape which had to be analysed on a digital computer. But in 1962 when test observations were expected, the Hamburg University computer *TR4* would not be ready for another one or two years, especially to read paper tape. An agreement was therefore obtained with the institute of Professor Unger in Bonn to use the *ER56* computer for free, and funds were obtained for 15 travels to Bonn. This solution would have required provisional programming in a special language different from ALGOL. Luckily this was never needed.

In 1963 a suitable computer became available at the Copenhagen Observatory, and I used it at a short visit about 18 July 1963, assisted by the young student Richard West whom Professor Anders Reiz had introduced to me.

It became clear to me that this computer would be the best for our purpose. I also considered computers from Zuse and IBM, but none of them could match the GIER with respect to speed or handling of punched tape or to the ease of programming with the ALGOL language.

A plan found in an application for funding of 28 August 1963 foresees again a 2-year observation period as the plan in 1961, but beginning in 1965, total cost of 1.6 million DM, including salaries, a new pavilion in Perth and an IBM 1620 model-2 computer.

In reality, the observations began in late 1967, took 5 years and a Danish GIER computer was used. GIER (see GIER (1961) and the following section) was 20 per cent cheaper and 20 times faster than the *IBM 1620* in computing and the tape reading and punching were ten times faster; the *IBM 1620* (see IBM 1959 and 1962) would have meant disaster for the expedition.

I remember Gerhard Holst telling me that Haffner wanted an IBM 1620 and that I said I would leave the project if that happened. Later on in Australia, it turned out that even with the GIER it was just about possible to keep up with the reduction of all the observations.

I fully understand that the ideas of Haffner and von der Heide had to go towards a computer from a big firm like IBM which could presumably offer service as far away as Australia. That seemed impossible for a small Danish firm, but *Regnecentralen* did it for GIER! But why? The contract must have cost them a lot of money, placing a full time paid engineer in Australia all the years. Did they do it for the possible publicity? Or for the sake of astronomy?

We do not know how their decision was taken in this case, but I have recently asked the professors Peter Naur and Christian Gram who worked in or with the Regnecentralen in those years, though not in the marketing area. They said that it was a general policy of the firm to aim for universities, thus reaching also the new generation of users, the students. Selling computers abroad and to renowned institutes counted much for the firm.

**The GIER computer**
GIER is an acronym for *"Geodætisk Instituts Elektroniske Regnemaskine"* (Institute of Geodetics Electronic Calculator) and was introduced there on September 14, 1961. The Copenhagen Observatory had a GIER installed in September 1962, at first only with machine programming, but soon with an ALGOL compiler.

The GIER computer, in the brown cupboard of Figure 15, was one of the first fully transistorized computers in the world. It had the following capacities RAM: 0.000,005 Gbytes; Drum: 0.000,07 Gbytes; CPU: 0.000,0007 Gflops, i.e. more than one million times less than any laptop computer of the year 2012. The external storage was on 8-channel punched tape, one role of tape could contain 0.00012 Gbytes. A tape could be read at the speed of 2000 characters per second which was unmatched in those years.

The GIER was about ten times faster in tape input and output and in computing than the IBM 1620 which had been considered in the above mentioned letter of 28 August 1963.

A small group from Bergedorf visited an institute in Heidelberg where a GIER was installed and they went to Denmark, without my prior knowledge, to check my important decision about the GIER computer. It was confirmed.

GIER was delivered in Bergedorf on a lorry in the afternoon of 2 November 1964 and became operational the same evening. We had the luck that a more powerful ALGOL compiler had been released shortly before our computer was delivered. GIER soon became the computer of the whole observatory, easily accessible as it was on location at the first floor, above the office of the director. It was a much better option than to use the computer at the university. It was used for 2.5 years during 9300 hours for observatory purposes before it was moved to Perth.

The annual report for 1964 states as follows:

*"Im Rahmen des SRS-Programms steht schliesslich die Beschaffung eines Elektronenrechners, der für die Dauer der Expedition nach Perth mitgenommen werden soll. In Zusammenarbeit mit der Deutschen Forschungsgemeinschaft wurde mit den Firmen Eurocomp, IBM, Zuse, Remington, CDC und Regnecentralen über eine für das SRS-Programm geeignete nicht zu teure Rechenanlage verhandelt. Die Entscheidung fiel auf die Rechenanlage GIER der Regnecentralen, nachdem Herr Holst im Mai bei der Gelegenheit einer Besprechung über einige Einzelfragen des SRS-Programms im ARI in Heidelberg eine beim dortigen Max-Planck Institut aufgestellte GIER-Anlage besichtigen konnte."*

The Danish manufacturer, *Regnecentralen*, agreed to send an engineer full time to Perth for the duration of the expedition to guarantee immediate repair if needed. The firm must have chosen their best ever GIER because repair was only needed a few times during the five years, so the engineer had nearly full time for himself. He stayed on in Australia with his family also after the expedition. GIER remained operational until at least 1980, and I used it at my last visit in February 1980.

Von der Heide had expected that the engineer Jørgen Sørensen would take part in the expedition also as observer, but that turned out to be a misunderstanding. I have not seen the contract where that should have been stated. But Sørensen took much care of the electronics for the meridian circle although he sometimes was hampered by severe health problems.

## 6. Building the instrument

**Seven years working on the ideas: 1960-67**

During the years of implementation I worked on the electronics, design of the mechanics, programming and testing of components and system, assisted especially by Herrn Holst and Herrn Ziegler; I am keeping here the way we spoke to each other in Germany, and we always spoke German, of course using the polite "Sie" for you; their first names were Gerhard and Ulrich, respectively, but they were never used until many years later, and I was Herr Høg to them.

There were six locations in the observatory area for my activities: Meridian circle building, main building with the GIER computer and the library, the Sonnenbau with my electronics workshop, the Lippert with my office, and the Schmidt for talking with colleagues. I used my Danish bicycle all the time which was unusual in Germany and some have mentioned a funny sight: Herr Høg often standing in the pedals to speed up, they grinned.

But how could the telephone operator reach me when I was called by somebody about an electronic component? I could not always call her and tell where I was going, but she was quite good until the job was taken over by the secretary who also had other things to look after. Once, I proposed to install a calling system with a cable around the observatory area. Such systems had then become commercially available. But that was too much for the director, and who should pay? A colleague consoled me saying that I was of course right in principle, but that was too advanced for the time.

My employment in Hamburg was temporary to begin with, but I did not worry because I am fundamentally optimistic and I was very occupied by my work. I heard rumours that the director tried to get a permanent position especially for one with my capabilities. I did not believe the rumours, but in 1964 I became Deutscher Beamter auf Lebenszeit and was sworn in with others in front of the Hamburger Senat in the 19th century-Renaissance city hall. In 1970 I became Hauptobservator and have received a pension from Germany since 1997 when I was 65. At the same time I stepped down in my Danish salary and was freed for obligations to teach.

My presence in Hamburg runs as follows: 1 October 1958 - Entrance as Stipendiat, the first 10 month funded by Deutsche Akademische Austauschdient then by a NATO Science Fellowship; 1961 - *wissenschaftlicher Angestellter (Deutsche Forschungs Gemeinschaft)*; 1 Januar 1964 - Observator; 23 April 1970 - *Hauptobservator*; 31 August 1973 - End of imployment.

**Building and testing**

These years have been reconstructed from memory and by means of the Jahresberichte der Hamburger Sternwarte found in a hard-cover book, HS (1973). The work took much longer than expected for several reasons: we had generally underestimated the size of the project; assistance for the electronics was not imployed in time as mentioned above; industry did not always deliver in time. But when I look back I must say that I enjoyed the work with many good collaborators.

In 1960 the AEG in Berlin was finally selected to deliver the electronics and I met the engineers there several times. I always used the opportunity to visit East Berlin, e.g. the famous Brecht theatre at *Schiffbauerdamm*. I saw and heard Helene Weigel as *Mutter Courage* and was fascinated by Kurt Weill's music, I saw *"Der gute Mensch von Sezuan"*.

The glass plate with finely etched slits for the star crossing was manufactured by the Heidenhain firm in Bavaria which I also visited. But I was never allowed to see how they manufactured such a plate, the process being of great commercial and military importance.

The Jahresbericht for 1961 describes the planned instrument and the expected performance. Delivery is expected in early 1962 and tests will be carried out at the 90 mm broken transit instrument from Askania.

**First testing in 1962 – with Schmidt-Kaler**

In 1962 such tests were completed and showed the expected limiting magnitude of 9. But the digital electronics, the *Zähllocher* in Figure 10, showed errors in the construction and was returned to AEG in Berlin for repair. The slit plate was measured and was not sufficiently accurate (HS 1969, letter to DFG of 10 July 1962). It was replaced by a new plate from the firm.

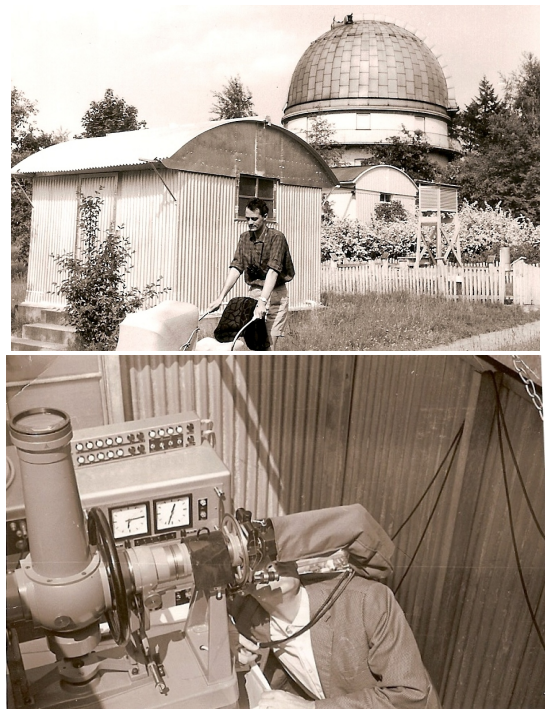

**Figure 9** The pavilion with the broken transit instrument in 1964, the author in front and the big Schmidt behind. Below is a picture from the testing with this instrument in 1962. - EH

The instrument was located in a small pavilion which no longer exists, to the right of the walking path to the Schmidt, see Figure 9. The first night I locked the door of the pavilion to avoid disturbance from any uninvited guests since I could not risk loosing any clear hour. This did not hold back a special guest, Th. Schmidt-Kaler who pulled the door in vain and then jumped in through the roof and asked if the door was locked because something of historic significance was taking place! Typically Schmidt-Kaler, the jump and the words, but we respected each other. I gave a talk on his invitation in the *Nordrhein-Westfälische Akademie der Wissenschaften* in Düsseldorf, and in 2005 I was guest with my wife in his house near Würzburg.

When the succession of Heckmann was discussed in those years at our coffee round in the Schmidt we could agree that we wanted someone like Blaauw or Voigt and we knew they would not come, because we were aware that the once famous *Hamburger Sternwarte* was not so attractive anymore. Of course we were not asked. We could also agree about two persons we did NOT want: Behr and Schmidt-Kaler. In effect, first Professor Hans Haffner took over the responsibility for the Perth expedition, followed by the professors G. Traving and later A. Behr.

**Will Perth Observatory survive 1962?**

The future of the Perth Observatory was very critical in the year 1962. This appears from a letter from Hans Haffner to Bart J. Bok, director of Mount Stromlo Observatory of 6 November 1962 and from Bok's answer of 13 November (in Høg 2012). I quote from Bok's letter: *"My advice is that you hold your fire until … and that then you let go promptly and firmly and tell the Western Australian Government that it would be criminal and sad if they were to close the Perth Observatory"*. He continues later: *"The Hamburg Meridian Circle will of course be welcome at Mount Stromlo Observatory if the Perth Observatory fails you, but I hope this will not come to be."*

In effect, the things went well in Perth thanks to concerted efforts of great people to make astrometry move forward! In those years I personally heard only little from these fights which now appear to me very lively in the correspondence (Høg 2012).

Von der Heide writes on 8 November to Professor Dirk Brouwer, director of the Yale University Observatory on behalf of Heckmann and Haffner asking for support. He mentions that the DFG has recently looked favourably at a cost estimate of 2000,000 DM (500,000 USD). Brouwer gives his support as G.M. Clemence, director of U.S. Naval Observatory had already done.

Perth Observatory survived closure in 1962 by moving into fine new buildings in the hills to the east of Perth (Figures 14, 15 and 17) in late 1965 and had a total staff of around 15 in 1970 of whom 3 were astronomers, in addition came the eight observers from Hamburg. By 1972 the number of astronomers rose to 5.

Craig Bowers wrote to me: *"... in 1987 the then Government of the day cut the staff numbers in half, to a level that I think was below critical mass for operation of a scientific program, however they have soldiered on."*

In February 2013 Bowers adds: The Department of Environment and Conservation, the governing body for the Perth Observatory commented *"....the focus in astronomical research is shifting away from the work traditionally done at the observatory, towards the more modern methods being delivered through the Square Kilometer Array technology."*[2]

The WA environment minister, Mr Marmion, commented on "Research at Perth Observatory axed" by saying *"...optical astronomy was simply being overtaken by radioastronomy...."*[3]

---

[2] Retrieved from The ABC News: http://www.abc.net.au/news/2013-01-22/research-cut-at-the-perth-observatory/4478756.

[3] Retrieved from SBS World News Australia: http://www.sbs.com.au/news/article/1729618/WA-minister-defends-observatory-decision

**Construction goes on in Bergedorf**

In 1963 eight micrometers (Figures 12 and 14) for reading the declination circles were being built. The two circles obtained new gold covers and division lines in the workshop of the US Naval Observatory in Washington DC. A row of dots was also provided for automatic recording of degrees and minutes, see Figure 20, but I do not recall that a reading was ever implemented.

The pivots (see Figures 18 and 19) were being renewed at the firm Heidenreich und Harbeck, Hamburg, using special steel. The final grinding and testing will be done by von der Heide. - The completed meridian observations of the AGK3R stars were sent to USNO by von der Heide.

In the months May to July Holst and Ziegler observe with the *"Høg-micrometer"* on the meridian circle. These observations shall be evaluated in order to define an optimal slit plate and to determine the accuracy of the observations - and this was done at the GIER computer in Copenhagen. Wolfhard Schlosser built a program unit for the digital clock for the micrometer.

The plan is to move the instrument to Perth in 1965 for a period of two years. The observations will be punched on tapes and be analysed on an *IBM 1620* computer – but it came differently.

In 1964 the eight micrometers have been completed and the plan for the main micrometer is nearly finished so that the manufacturing can begin shortly. Von der Heide has completed the grinding and testing of the pivots.

The funding for the purchase of GIER and the expedition for 2.5 years with six observers has been granted by the *Stiftung Volkswagenwerk*.

According to a report to DFG of 9 December 1965 (HS 1969) the work had progressed as follows. The second *Zähllocher* was delivered by AEG in February 1965. Von der Heide had polished the divided circles because they had rather coarse defects. The slit plates for the micrometers were ordered from Heidenhain in April but were not yet fully delivered – firms were often late on delivery. Von der Heide and Høg had designed the main micrometer and it was being built in the institute workshop. The electronic control was built by Dipl. Phys. F. von Fischer-Treuenfeld imployed since 1 October 1964 and he stayed until 31 December 1967. Von Fischer-Treuenfeld also made changes in the two *Zähllocher* (Figure 10) so that they could be used in a redundant mode if one of them should fail – this proved to be very important in Australia in 1970 as reported below. This mode was already proposed in the letter of 9 May 1961 from von der Heide to Heckmann.

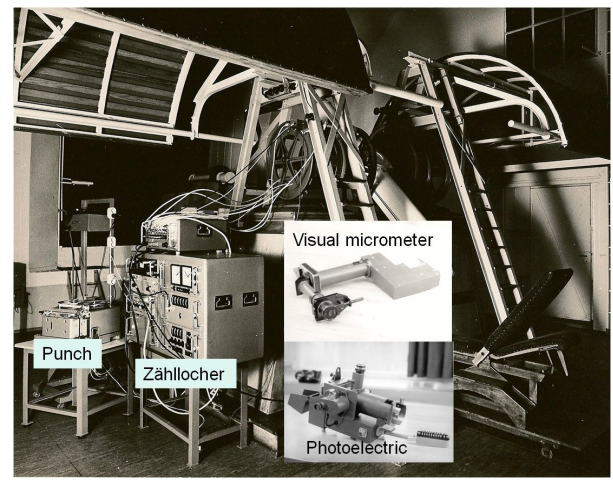

**Figure 10** Hamburg 1964, the electronics for reading the declination circle. Left: the paper tape punch and the electronics box *Zähllocher*, an identical box in the hut was used for the star observation. Inserts: a visual micrometer, and a below: a photoelectric micrometer for the star. A photo of the micrometer for circle reading has not been found. – WD

In 1966 the division lines on the circles had been brought in good condition by von der Heide and the illumination of the lines had been improved. Von Fischer-Treuenfeld could then begin the series of measurements required for the determination of the division corrections with the general symmetric method (Høg 1961b). He completed the work in 1967 which was published as dissertation, von Fischer-Treuenfeld (1968).

Further test observations were obtained in 1967 and were analysed before the instrument and the GIER computer were packed for shipment.

**Final instrumentation for Perth**

The final meridian circle instrumentation of 1966 is shown in Figures 11 and 12, ready for being dismantled and sent to Perth. The photoelectric micrometer is attached at the lower end of the telescope. The micrometer is painted brown. The cable on the floor (Figure 12) leads to a small push bottom unit, here placed on the ladder. It was held by the operator standing in dark during the night. He pushed the bottom after setting the telescope in declination.

Behind the ladder appear some of the cables from the electronics (Figure 12) to the barely visible circle micrometers at the far end of the long microscope tubes. A circle microscope consists of a microscope tube and an illumination tube which is better seen at right in Figure 14.

Through the open door at left you glimpse into the hut where the night assistant was sitting at light so that he, or usually she, could read the star list and tell the operator the declination of the next star and set various switches on the electronics in front of her.

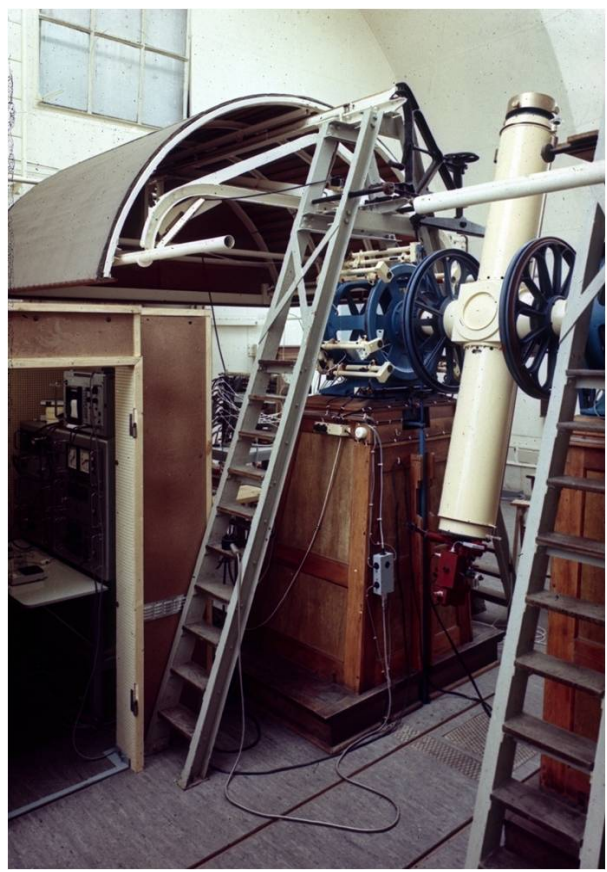

**Figure 12** Hamburg 1966 - The Repsold meridian circle ready for Perth. The observer set the telescope to the declination as ordered by the assistant sitting with the star lists in the hut at left. He started the recording when he saw the star at the proper place in the field of view. – WD

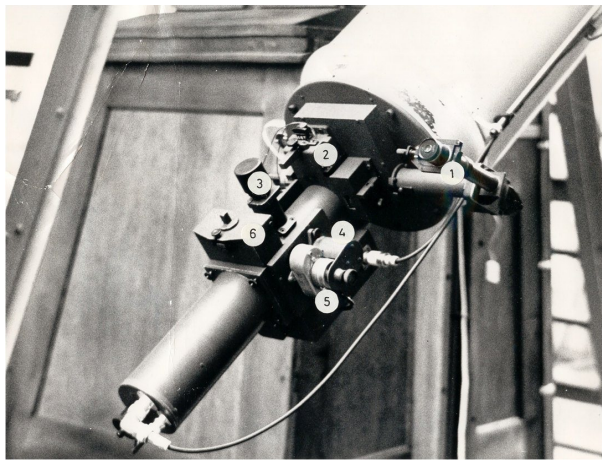

**Figure 11** Hamburg Bergedorf – The first slit micrometer on a meridian circle 1966. The instrument was semi automatic with manual setting of the telescope and digital recording on 8-channel punched tapes of the star observation and of the circle readings. - WD

## 7. Three ideas

I recall from these years a flustering that Høg is more interested in instrumentation than in astronomy, e.g. in observations, and he does not publish anything. My elder colleague in the Lippert was Professor Arno Wachmann and he was a very good observer, producing light curves in the hundreds or even thousands from patient observations over many years. That is one kind of useful astronomy. I was engaged in another kind of activity, to create better tools for astrometry and I could see they would be used for observations in Perth. *I always just wanted to do* *useful work for astronomy* and I had the great luck – all my life - to work with many people who had that same aim.

I wrote scientific papers about the progress with the instrumentation and about the theoretical basis for astrometry in general and for the photo-electric multislit micrometer as I called it, taking into account all error sources from the atmosphere, from photon statistics etc. (Høg 1968, 1970, 1972).

**Double star micrometer and Chr. de Vegt**

Three ideas beyond the Perth expedition were in my mind, one was to observe double stars with a scanning micrometer and photometer to be built for the big refractor. In 1965 I obtained funding from the DFG after test observations with counters and the meridian circle slit system on the refractor had shown good results. The system (Høg 1971b) was implemented with an on-line computer, the first in Bergedorf. It was the small rack-mounted 12-bit *PDP-8/S* computer. *PDP-8* was introduced by *Digital Equipment Corporation* (DEC) in 1965 and the first commercially available mini-computer. I used the scanner on double stars and on planetary nebulae and on a comet in collaboration with L. Kohoutek, but I never had time to publish the observations.

The photometer was used for timing of the optical Crab pulsar with the big refractor (Høg & Lohsen 1970), the first such timings in Europe after the pulsations had been discovered in 1969. The pulsations were followed for five years by my student Herr Lohsen and the whole idea to observe the pulsar was due to him. This work was possible because the photometer at the big refractor used photon counting techniques with accurate timing in UT, similarly to the system at the meridian circle.

The photometer was also used by Christian de Vegt for determination of six diameters of stars from observation of Moon occultations during the period 1969-75, see de Vegt (1976). It is beyond the scope of this report to cover any other astrometry in Hamburg than that at the meridian circle. But I must mention some of the very important work on astrometry by my younger colleague Professor Christian de Vegt (1936-2002), a work completely independent of mine and over many more years in Hamburg.

De Vegt and his collaborators developed the block-adjustment procedures to a practical tool, allowing astrograph-type observations to tie together for a global solution, similar to what Hipparcos did later. De Vegt acquired a new telescope in the million DM range which was quite an accomplishment to a place like Hamburg, and this astrograph did extremely accurate astrometry, for small angles on the sky better than any meridian circle. De Vegt began the extragalactic link program for Hipparcos in the 1970s in Hamburg with observations of ICRF sources. He also pioneered CCD observing for astrometry leading the way into future applications in this area and built accurate plate measure machines to allow complete extraction of historic astrometric data, particularly valuable for proper motions of stars fainter than the primary reference frame stars. The SAO facility retrieves 196 abstracts for C. de Vegt from the years 1966 to 2004.

**New type of meridian circle**

The other idea of mine was to develop a new type of meridian circle, the glass meridian circle (Høg 1971a, 1973, 1974a and b). The use of a horizontal telescope (Figure 13) was supposed to solve long standing problems of flexure and refraction in fundamental astrometry. In fact, it led to the discovery much later of a hitherto unknown refraction inside meridian circle telescopes and to the elimination of this effect by a simple ventilation of the telescope tube, Høg & Fabricius (1988). This internal refraction could be as large as 1 arcsec and it was quite variable during the night. It has badly affected all determinations of the external, atmospheric refraction in the history of meridian observations. Through the removal of internal refraction a much better determination of atmospheric refraction became possible, since the two effects are inseparably confounded, see Fig. 8 in the paper by Høg & Fabricius (Claus Fabricius *1954)

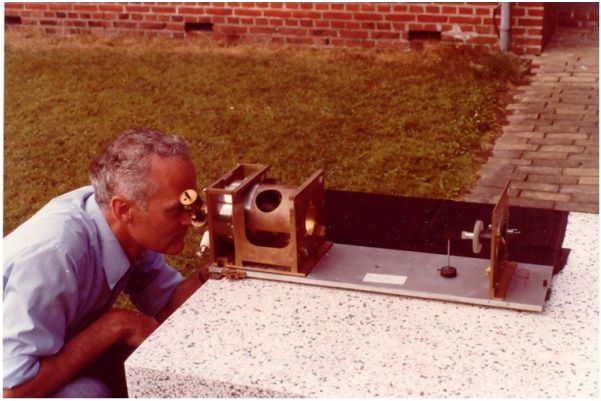

**Figure 13** A new type of meridian circle, the glass meridian circle, was proposed in 1971, here a model from about 1973. - EH

Work on a glass meridian circle was pursued in the 1980s in collaboration with Chinese astronomers, but after the success of Hipparcos the meridian circle became an obsolete type of astrometric instrument. It has been the fundamental instrument for measurement of large angles on the sky for 200 years, but with the *Tycho-2 Catalogue* (Høg et al. 2000) available containing 2.5 million astrometric stars only relative observations were needed, i.e. only small angles within a few degrees, and in addition this gives higher accuracy. Such observations can easily be obtained with a large CCD in an astrometric telescope which can be pointed to any area of the sky whereas the meridian circle is limited to the meridian. Nevertheless, some meridian circles have remained very productive with CCD detectors as described in the preceeding sections on "Photon-counting astrometry".

**Automatic astrometry**

The third idea was to build a fully automatic meridian circle which was realized on La Palma as mentioned above. To go even further and develop a satellite for astrometry was a vision of French astronomers which began in 1964 according to Kovalevsky (2009) and which was pursued especially by Lacroute (1967, 1974), but it was beyond my mental horizon until I was invited by ESA in 1975 (Høg 2011a).

# 8. Australia

**Hamburg expedition in Perth 1967-1972**

The following reconstruction of the five years is made from my memory after more than 40 years and by means of the archive of the Perth expedition, Høg (2012), including letters between 14 February 1960 and the end of 1971. But at present I am not writing the history of the expedition in the detail that a historian might wish to do and I will therefore mainly describe my own work. A historical review of the expedition is given by Behr (1976).

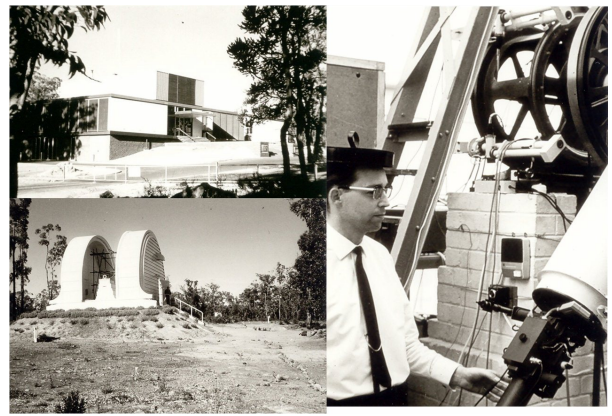

**Figure 14** Perth Observatory 1967. Left: Main building and the meridian dome. Right: Gerhard Holst (with the dewcap!) at the meridian circle. - IH

The ship with the instrument arrived in good condition in Perth. The participants had arrived before the instrument and arranged for housing etc. They set up the instrument and observations were begun in November 1967 (Figures 14 and 15). I went to the IAU General Assemby in Prague in August of that year and was anxious whether there were problems with the instrument which was quite critical at some points, especially with the nadir observations via the mercury pool. But my clever wife as usual advised me: nobody in Prague would know about any problems, so just be confident! I did mention my worry to my friend G. van Herk when I saw him in Prague, but he also calmed me.

The preparations for the expedition to Perth have been reported as I saw and worked on them. I did not take part in the expedition itself except during three visits to Perth of a few weeks duration in order to help out on various issues. Gerhard Holst was my direct support during these visits as he understood the instrument with the mechanics and electronics better than anyone, and during the whole expedition he was therefore a central person.

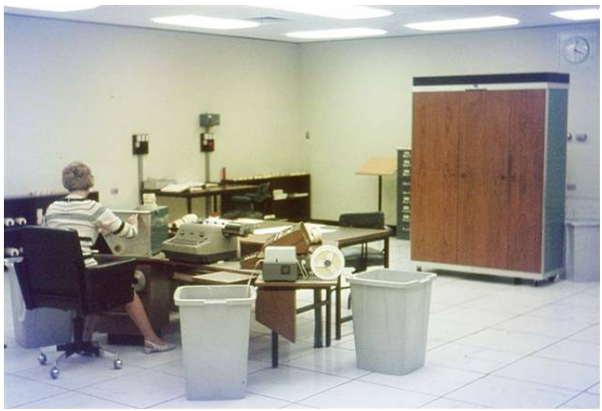

**Figure 15**  Computer room of Perth Observatory 1971. GIER in the brown cupboard was one of the first transistorized computers. Mrs Ilse Holst at the reader for 8-channel punched tape, one role could contain 0.00012 Gbytes. The tape from reader or punch poured into a large basket and could then be quickly rolled up. - BL

**Visits to Perth in 1969**

Professor Alfred Behr in Bergedorf had from 1 January 1968 taken over the responsibility with respect to the funding from Stiftung *Volkswagenwerk*, thus succeeding Professor G. Traving who moved to Heidelberg. This responsibility included the correspondence with the leader in Australia, Dr. Johann von der Heide, especially in administrative matters, and I did not have access to this correspondence kept by Behr. Only when I asked Behr, and I sometimes had to insist, did I get a limited visibility, limited for reasons I could guess about. But I always felt concerned about the instrument, the observations and the hard working people in Australia as if I were scientifically responsible and my correspondence with von der Heide and Holst about problems with the data reduction and electronics is extensive.

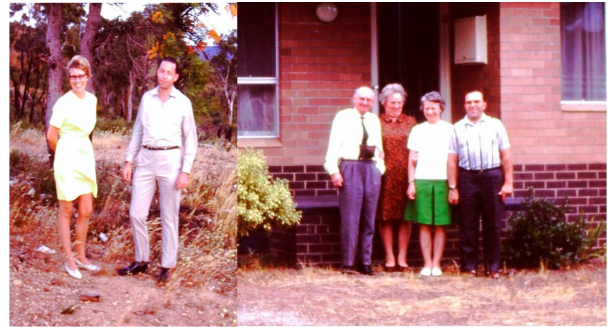

**Figure 16**  The six observers from Hamburg in 1969. From left: Ilse and Gerhard Holst, Johann and Helene von der Heide, Martha and Ulrich Ziegler. - EH

A long letter of 25 July 1968 from von der Heide speaks of many issues and closes by *"again thanking me for all the work and time I have spent in modernising the meridian circle."* He also asks if my new double-star micrometer is performing well. I am happy to find this letter again in December 2012 after all those years because later on in 1969 his attitude to me is very different.

On 4 November 1968 von der Heide reports to Professor Walter Fricke about the status of the observations and asks for an extension by 2.5 years from December 1969 because the observations are behind schedule. Fricke replies on 16 December by giving his support for an extension and by expressing his satisfaction with the quality and quantity of what has been achieved. Alfred Behr and Walter Fricke visit Perth in June 1969 in preparation of further decisions, including application for funding. The work load on the observers was so high that extra observers were needed. A visit by me was included in the plan and I arrived in Perth on 13 November and stayed a few weeks.

On the first day in the observatory I met the six observers to discuss issues especially about the GIER programs. At lunch I remember how Mrs Helene von der Heide (1910-2006) enjoyed seeing photos of my family with the two children, she was always like a good mother to everyone around, but we remained of course Herr Høg and Frau von der Heide to each other.

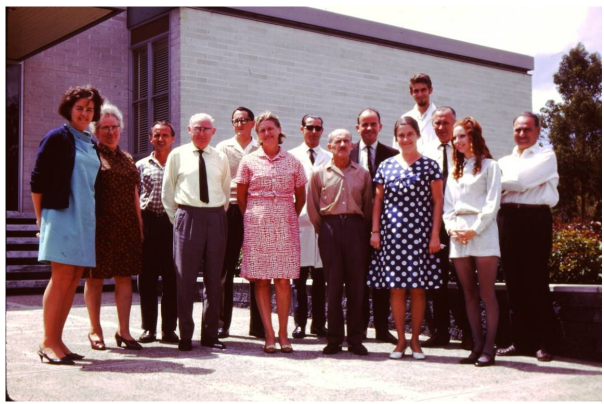

**Figure 17** Members of the Perth Observatory staff in December 1969. From left: Miss I. Wardrop, Mrs H. von der Heide, Mr J. Pratt, Mr J. von der Heide, Mr L. Cloud, Mrs K. Pratt, Mr D. Gans, Mr R. Hutchins, Mr M. Candy, H. Sharpe, Mr D. Harwood, Mr. D. Griffiths, S. Constantine, Mr J. Harris. – EH

After lunch I was alone with Gerhard Holst and we computed "clamp differences"; the instrument can be lifted by a crane, be rotated 180 degrees, and be placed in the other bearings. Observations of the same star under these two conditions should in principle give the same position. The differences allow important checks of the instrument as is described in full detail by Høg (1976: 26ff). In the evening in the hotel I made diagrams to present to the observers the next day.

The diagrams showed that the right ascensions were OK, but not at all the declinations. A very large discrepancy along the meridian of 2 seconds of arc was discovered, much larger than the mean error of 0.30 seconds of arc for the declinations in the final catalogue. This could therefore give large systematic errors in the final catalogue thus decreasing its value in improving the fundamental catalogue FK4, an undertaking Walter Fricke and his collaborators in Heidelberg completed in 1988 with the publication of FK5. With this in mind I used all my energy to find an explanation, but we faced a mystery for more than two months, during my stay in Perth and after my return to Hamburg, until a very simple instrumental defect was found and completely repaired.

When von der Heide came to the observatory the next day and saw that I had computed the clamp differences he became quite upset. I did not have his permission to do so and he had intended to do that himself, but the observing and other duties had not permitted, an argument he repeated in letters. Before I went to Perth I had discussions with Behr where he offered to give me a letter to show to von der Heide if needed, arguing that I should determine the clamp differences. But I did not want such a letter because I believed von der Heide would be happy if I did it.

Von der Heide was upset, but I did not feel that during the rest of my visit. He took me on a nice picknick into the wild Australian bush with his wife and the two Zieglers and I have a film taken on super-8, now on CD, where he cooks water for tea on an open fire at the roadside. Ziegler performs on the film in his jolly mood, he was always spreading good humour around him. He had joined the Rotary Club in Perth on an invitation which von der Heide had declined. Mrs Ziegler was more reserved, in fact she did not like the time in Perth and she suffered from the long five years.

Besides of the work in the observatory, hunting especially for an explanation to the declination problem I was entertained in many ways. Dan Trout, the electronics engineer, took me to his home where I met his wife and, surprisingly, my old friend Benny Clock from the US Naval Observatory who was going around the world to synchronize clocks to microsecond with his "travelling quartz clock". Dan also took me on sight-seeing in Perth. Ilse and Gerhard Holst took me to a park north of Perth were we saw Koalas. Sørensens and Nikoloffs arranged a great party one day and evening in their homes with many guests and enormous amounts of food and drinks; von der Heides excused themselves for this happening. I went to the beach with Miss Wardrop and a third person and we were nearly carried away by a big wave into the Indian Ocean. I was invited to Ulrich Ziegler, to Jørgen Sørensen, to John Harris and to Michael Candy, and an evening I went with Mrs Candy and Miss Wardrop to see the famous absurd play *"Waiting for Godot"*.

**The micrometer was tilting !**

Back in Bergedorf, I discuss the declination problem with Behr and others and propose various possible explanations in my report of 23 December. I also emphasize the high quality of the observations and that the total performance hitherto exceeds any other meridian circle by a factor of four. On 26/27 January 1970 I visit Fricke in Heidelberg to discuss the situation, he is also concerned about the large clamp differences.

In the mean time von der Heide studies the instrument in Perth in order to locate the source of the problem. But in a letter to Behr of 16 January he complains about the many different ideas coming from me, and on 5 February Behr supports his view that *"Høg is a difficult person, changing his ideas from one day to the next, but we have to get along with him"*.

On 29 January I am able to write to von der Heide precisely what should be done. I had discussed with Behr and especially with Herr Schultz in the Bergedorf workshop, could it be that the micrometer was not well enough fixed to the telescope tube?? The micrometer could then tilt a bit at a certain declination resulting in the observed clamp differences. Herr Schultz had no drawings of the instrument, this was normal in Bergedorf, but he remembered how the micrometer was fixed by a screw and this is shown in Figure 4 of Høg (1976).

On 12 February I explain to von der Heide that my many letters to him were meant to replace the conversations one would normally have and that I did not mean to give him orders, but merely suggestions. I close by saying that he at the location may have other arguments than those known to me.

On 20 February von der Heide reports in all detail to Behr that his visual inspection had now shown that in fact the fixing screw did not do what it should. He will have it repaired and observations should soon show whether the error was remedied. He closes the letter asking Behr to inform me and to pass his cordial regards to me.

Soon after, we received information about the satisfactory operation. In consequence, I distributed a report dated 14 October to the six observatories with a Repsold meridian circle, with copy to von der Heide, warning them about the serious defect in the fixation of the micrometer. It is suspected that this defect could have caused errors in the observations with these instruments ever since they were set up, as is indicated by a flexure term of 1.1 second of arc in observations since 1913, see Dolberg et al. (1937: 113).

**Tension goes up and down**

There was other serious tension in these months. In the letter of 12 February 1970 I ask von der Heide to comment on a manuscript - if he can find the time to do so. Behr helped me to express clearly who had contributed to the work and the paper appeared as Høg (1972). In a long handwritten letter to Behr of 15 April von der Heide notes that I have not agreed to be an observer in Perth as Behr had proposed; I could not do that because it was my duty to observe with my newly built micrometer for double stars. Von der Heide blames me for escaping the hard observing work and for coming just a couple of weeks *"taking their time and pleasure of work"*. He had received a letter from Nemiro, president of the Commission for astrometry, where he saw that I had announced a talk about the photoelectric micrometer and he assumed I would present his, von der Heide's, observations. He himself was going to the IAU congress in Brighton in August and intended to give a presentation himself.

Behr showed me the letter which horrified me, I had not made any proposal to talk in Brighton.

But neither Behr nor Fricke seemed to believe me. On 8 May I explain in a letter to von der Heide that I had not had any correspondence with Nemiro at all, and that I will ask him for clarification. But already before this clarification comes from Nemiro, von der Heide encourages me in a letter of 19 June to present my work, including the systematic errors of FK4 and he looks forward to further talks with me in Brighton.

On 15 July I write to von der Heide that Nemiro had now explained to me that he had *"proposed a Provisional Agenda ... and mentioned probable speakers to the Organizing Committee (of which von der Heide was a member). Now the Agenda is approved, and I (Nemiro) ask you and Dr.J.E.B. von der Heide to make a report on the new Meridian Circle of the Hamburg Observatory."* I write to von der Heide that I will not present anything. This results in a letter from von der Heide of 7 August urging me to present the instrument and the observations up to December 1969. He will himself present the later observations. – All this illustrates that there was severe tension sometimes and that ALL involved persons were willing to calm the waves soon after.

**Eight months without observing**

From June 1970, the six Bergedorf observers had vacation in the sequence Ziegler, Holst, von der Heide. It was their first vacation in Germany after three years in Perth and before the last period should begin. In the extended period until December 1971 two new observers from Germany had joined the team since June 1970, Bernd Loibl (*1944) and Mrs Karin Loibl (*1941). Four observers from Australia participated until August 1972 in various periods, Dr I. Nikoloff beginning already in March 1969 and Miss I. Wardrop in January 1970, see Høg, von der Heide et al. (1976: 20).

Von der Heide's presence in Bergedorf in September 1970 was used for a meeting of Behr, Dieckvoss, Gliese, Haug, von der Heide, Høg, Neckel, and Wehmeyer about questions of the observing program, but with no occasion to ask questions or to speak about technical problems. But there was a very serious problem: they had not been able to observe at all since 15 June because of electronic problems when replacing the Lorenz punch with a Facit for the declination readings.

I discovered only much later, in February 1971, that something was wrong and asked Behr. He said he believed I knew, but I did not. I requested to see the correspondence which Behr reluctantly showed me. Then I asked him to call Weigert, Traving and the whole scientific staff to a meeting which took place in the Schmidt.

After some questions and explanations from Behr, I said that they could easily start again and observe with reduced speed. The normal operation was that the star observations were handled and punched by one *Zähllocher* and the declination measurement and punching by another, each with its own punch. The electronic problem was that the *Zähllocher* for the declination could not be used because the change to Facit punch required a new electronics to be built and this had not been accomplished. My advice was to turn a switch we had installed on the back side of both *Zähllocher* for the sake of redundancy. If the switch would be turned all measurements from star and from the circle would go to the *Zähllocher* with the *Facit* punch. This mode was proposed in the letter of 9 May 1961 from von der Heide to Heckmann.

My proposal was met with some astonishment, but a message (telegram) was sent to Perth, and in fact it worked, they could again begin observation with reduced speed on 1 March. But the electronic problem had to be solved, therefore Behr and I departed for Perth on 2 March 1971 to help, for both us it was the second visit. Behr stayed a few days, I three weeks, and again we felt very welcome, a big party was arranged for us. But the electronic problem was not solved during our visit because the available expertise was insufficient.

A drama happened when I was there, one of the most involved persons went away by car one day for so long and under such circumstances that we feared for his life, but he did come back.

Jørgen Sørensen was in charge of the electronics, much supported by Dan Trout, the electronics engineer of the Perth Observatory, by Tom Berg from the university, and by Holst and Ziegler whenever needed.

Many repairs of the observing equipment were required especially during the last part of the expedition, often quite difficult as when a transistor was broken and it turned out to be not available, nor its successor. A letter on three pages from April 1968, but undated, from our electronics engineer in Bergedorf, Herr Fritz Harde, to Herr Holst illuminates the huge problems with exchange of transistors very well. It is therefore astounding that the meridian circle could be kept operational during 20 years until 1987 by the Perth Observatory staff, Harwood (1990), and the GIER computer until at least 1980 for production of the Perth75 Catalogue by Nikoloff, Høg & Ayers (1982). In electronics design of the 1960s transistors were replaced by the standardized and much more reliable integrated circuits, but the expedition in Perth had to live and fight with transistor-based electronics. Many components had to be purchased in Hamburg and then sent to Perth.

**Collaboration with Behr and von der Heide**

I consider the repeated problems between von der Heide and me in the time after my first visit in December 1969 to be due to the heavy load of work and responsibility on him in Perth and to a sometimes failing health. Professor Behr was aware of the problem and was perhaps afraid that von der Heide could collapse which would have endangered the whole expedition. This could explain Behr's actions, including his lack of support for me at some occasions.

Clearly, if Behr from the beginning had kept me informed of technical problems in Perth I could have helped sooner. He did inform me much better after our common visit to Australia, which ran in a very positive spirit. The Germans and Australians received us very well and gave a party I still remember, but to tell about that would be too much in this context.

I should however add something: In 2005 my wife and I were visiting Hans-Heinrich Voigt in Göttingen and I used this opportunity together with Voigt to see Behr who lived alone in his house in a nearby village. Behr was very glad to see us and to hear of the great progress astrometry had made with Hipparcos. When I excused for taking perhaps too much of his time he smiled: *"You can't imagine how much time an old man has!"*

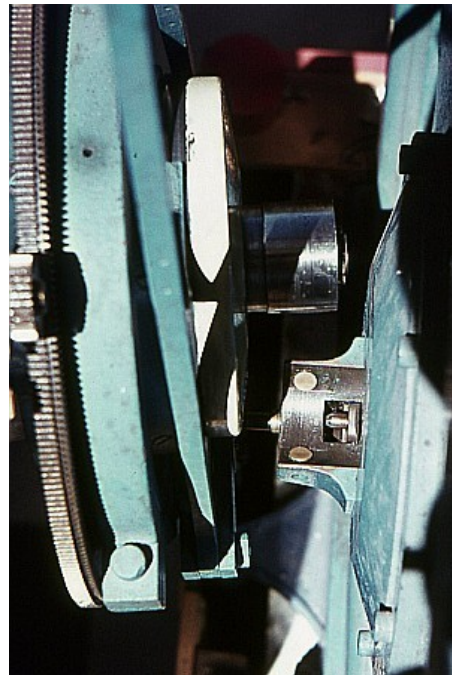

**Figure 18** The cylindrical pivot of steel. It glides on the two oiled circular bronze pads seen below. – BL

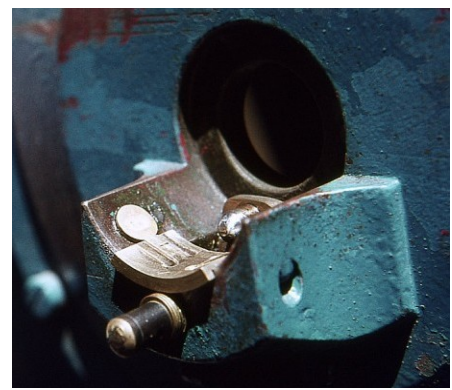

**Figure 19** The pivot bearing in a close view. - BL

During the visit to Perth in March 1971, Behr and I met the two new observers from Germany, a young student Bernd Loibl and his wife. Loibl invented and applied a new method to determine the errors of the pivots (Loibl 1978). The pivots (Figures 18, 19) form the ends of the east-west axis. They have a very accurately cylindrical form, but they still have errors which must be determined and be taken into account in the observations of the right ascensions. Loibl showed that the so far used methods gave incorrect results.

Support from Hamburg was needed and on 27 July 1971 von Fischer-Treuenfeld arrived in Perth for a stay of two months since he was very clever on electronics and on the special equipment of the expedition. With him present the installation of the Facit punch for circle reading was finally completed and observation could from about mid September go on at full speed again after more than one year with either complete stop or observation at reduced speed.

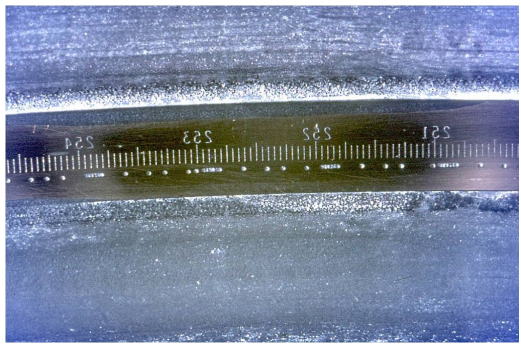

**Figure 20** Part of the Hamburg declination circle. An interval of 4 degrees is seen with 20 lines per degree. The dots would allow an automatic recording of degrees and minutes, which was however not implemented. – BL

On 16 September 1971 I joined the expedition in Perth for the third time to look after various issues. In view of the error found in the fixation of the micrometer, suspicion was directed towards all optical and mechanical parts of the meridian circle as described by Høg (1976: 34f). Von Fischer-Treuenfeld and I found that one of the four declination microscopes had been misplaced by 10 divisions, i.e. 30 minutes of arc (Figure 20), during all the five years of the Hamburg expedition. The offset microscope was moved to the correct position on 2 October 1971. It was therefore decided to make measurements for a new determination of all division line corrections for the catalogue completion. This was done by Holst and von Fischer-Treuenfeld during a visit to Perth in March 1974.

**Final catalogue made in Denmark**
The final Perth70 catalogue was produced after I had moved to Denmark in September 1973, but Gerhard Holst had suddenly quit his job in Bergedorf in January 1973, nine months before my return to Denmark because a position at the university computer centre was offered him and because it was clear that he could not rely on the observatory in the long run. His move came as a great shock to me since I had not been informed and the director in Bergedorf had permitted it without making any conditions with the computer centre that Holst should be allowed to help me although he was very important for the completion of the catalogue. I heard of this move from my wife when I returned from a travel and she says she has never seen me as pale as that, except when my mother phoned me in 1968 that my father had suddenly died.

My new colleague in Denmark, Leif Helmer did the computing in the Danish computing centre in Copenhagen and Holst gave all the support we needed, very conscientious and efficient as he always was. Altogether, in spite of many difficulties the expedition became a great success producing in 1976 the Perth70 catalogue (Høg, von der Heide et al. 1976) with accurate positions of 24,900 stars based on 110,000 individual meridian observations.

The catalogue stood any comparison at the time. Its place among the best astrometric catalogues during 200 years is shown in Table 1. The four last catalogues were obtained with the photon-counting astrometric technique invented in Hamburg in 1960 and developed for the expedition to Perth.

# 9. Review of Perth

I greatly admire the participants in the expedition (Figures 16 and 17) because they kept the instrument operational and made the computations during all the years, even two and a half years more than planned when they left Hamburg in 1967. The administration meant an additional work load especially for von der Heide and Holst.

I have previously (Høg 2011a and b) named seven persons from Bengt Strömgren to Ed van den Heuvel who acted in a chain in the modern development of astrometry leading from 1925 to the approval of Hipparcos in 1980. If anyone of these seven had been missing the development would not have taken place because there was no one else around who could have taken their role. I must note that also Johann von der Heide and Gerhard Holst were irreplaceable in this development.

It was hard time for all of them, but also a time which meant very much positive for their lives. Here I am quoting Ilse Holst from 2012 with whom Aase and I have kept the contact also after Gerhard had suddenly passed away in 2000. I always enjoyed my visits to Perth and felt I was welcome. I took often care of ordering spare parts helped by the administration in Bergedorf and I used a GIER computer in the Copenhagen Observatory to analyse tapes received from Perth when special problems required. I am grateful for the freedom I had in Bergedorf to care only about my research, including the Perth expedition, I had no other duties, neither in administration nor in teaching. Alfred Behr's administration and scientific advice were crucial for the success of the expedition and he deserves much credit for that.

**Tabel 1:** Astrometric performance 1800-2000. First class meridian circle catalogues and the Hipparcos and Tycho-2 catalogues are shown, the four last catalogues were obtained by photon-counting astrometry. The statistical weight of a catalogue is calculated as $W=N\sigma^{-2} 10^{-6}$ where $\sigma_{star}$ is the standard error of the position components in the catalogue as given by the observers.

| Publication Year | Name | Observation Years | N stars | $\sigma_{star}$ arcsec | Weight |
|---|---|---|---|---|---|
| 1814 | Piazzi | 21 | 7646 | 1.4 | 0.0034 |
| 1908 | Küstner | 10 | 10663 | 0.34 | 0.092 |
| 1952 | USNO | 8 | 5216 | 0.15 | 0.23 |
| 1976 | Perth70 | 5 | 24,900 | 0.17 | 0.86 |
| 1999 | Carlsberg CAMC | 14 | 181,000 | 0.06 | 50 |
| 1997 | Hipparcos | 3 | 118,000 | 0.001 | 120,000 |
| 2000 | Tycho-2 | 3 | 2.5 million | 0.06 | 700 |

With Johann von der Heide my relation was sometimes problematic as I have explained above and as I understood especially from Prof. Behr. Von der Heide did not want to take part in the catalogue production, Behr said. It was therefore with some anxiety that I phoned von der Heide in about 1978 and asked if I may visit him since I had the opportunity to make a stop in Hamburg. But to my great pleasure and relief he said: "Yes, of course". When I arrived, Johann von der Heide (1902-1995) was alone and had a lunch ready, asparagus with butter, which we enjoyed in a lively conversation. I now felt that an important part of my life had come to a happy conclusion.

He and his wife, who came home after a couple of hours, were obviously very happy to see me and to hear about the new plans for Hipparcos. At another visit shortly after I stayed overnight with them. In 1998 Aase and I visited Ilse and Gerhard Holst in Lohbrügge on our way back from the Gotha conference and I phoned Mrs Helene von der Heide (1910-2006). This was my last talk with her.

During the five years of the expedition also planets were recorded when they crossed especially long slits, so long that even Jupiter was inside. These recordings were kept on punched tape for later reduction after the expedition. In 1973 I met a young 23 years old student Lennart Lindegren (*1950) in the Lund Observatory and he took this reduction as task for his degree. The results for the five major planets Mars to Neptune and four minor planets were published (Lindegren & Høg 1977) and the accuracy was excellent thanks to Lindegren's analysis published also in 1977. He was a very good student, but when I introduced him to the new astrometry satellite project in 1976 his genius became obvious and without Lindegren there would clearly have been no Hipparcos satellite at all, see Høg (2008b).

The meridian observations were continued in Perth after the Hamburg expedition had ended in 1972. This resulted in the *Perth75 Catalogue* based on 60,000 individual observations of the 1156 FK4 and 1433 FK4 Supplement stars south of declination +38 degrees (Nikoloff, Høg & Ayers 1982). Nikoloff used the GIER computer for his catalogue and other purposes until at least 1980. After this time it was deconstructed probably in 1985, at the command of the then Government Astronomer/Director M Candy, according to information in 2012 from Craig Bowers, Honorary Historian to the Perth Observatory.

Dennis Harwood continued observations and published the "*Perth83: a Catalogue of Positions of 12,263 stars between 1980 to 1987*" (Harwood 1990) which he computed on a *Digital Equipment Corporation* (DEC) *PDP 11/10* computer after rewriting the ALGOL programs to FORTRAN.

The IAU Symposium No. 61, *New Problems in Astrometry,* was held in Perth from 13$^{th}$ to 17$^{th}$ August 1973 with 120 participants, under sponsorship of the IAU commissions 8, 24, 33 and 40 for respectively Positional Astronomy, Photographic Astrometry, Structure and Dynamics of the Galactic System and Radio Astronomy. Financial support was given by the Government of Western Australia. Proceedings appeared as Gliese, Murray, Tucker (1974).

## 10. My life with people

**Peter Naur**

Peter Naur (*1928) laid seeds to the great change of astrometry in two ways: he was my mentor from September 1953 and introduced me to astronomy, electronics and computing and secondly, his work on computer sciences was crucially important for the Perth expedition. The ALGOL programming language was developed in those years by a European-American group where Peter Naur played a leading role. ALGOL60 greatly facilitated my work on the computer programs for the Perth expedition since the symbolic programming was an immense step forward from the programming with numbers for the *IBM650* I used before.

Naur completely left astronomy in 1959 although he had been an astronomer since he was a boy. He became the first professor of computer sciences in Denmark and a guru of these matters. I visited him in 1978 for a talk and realized to my surprise that he was simply not following astronomy or the universe any more, and he meant it. In recent years he has again changed interest away from computing, writing e.g. *"An anatomy of human mental life"* in 2005. He won the 2005 Turing-award, also known as the *"Nobel-prize of computing science."*

Some time ago I saw a letter from the observatory leader of December 1953 when I had been given the task to work with the new meridian circle in Brorfelde. I was completely alone there and sometimes slept in a haystack when it became cloudy because there was no other building than just the pavilion. The letter quotes Naur for a warning that this task under such circumstances is likely to kill the interest in astronomy of a young man. But this did not happen with me.

Professor Bengt Strömgren had given me the task. I met him every summer when he returned from America and I reported regularly to him by letter about my observations. I admired him greatly and for me in those years a professor was someone like him – therefore I never had the ambition to become a professor, I could never become like him.

Naur was my mentor for three years and I became his assistant in Brorfelde, the new observatory 50 km to the west of Copenhagen. This lasted one year until I departed to Hamburg in September 1958. During that time I learnt to build electronics from the bottom based on a drawing, buying resistors, valves, capacitors etc. and soldering them for use at the new meridian circle. That knowledge was very useful in Bergedorf where it was something new.

Peter Naur has recently said in an interview with a Danish historian that he was fed up with the Copenhagen Observatory. There was no leadership present and the work with the meridian circle for which he was in charge was merely instrument building and then trivial observation of stars, one by one. It was not science, only like land-surveying on the sky.

I liked to work with him. He did mention his objections to me but not so strongly in those years, and it turned out that I had been put on the right track towards instrumentation and astrometry, exactly right for my personality and talents. I met him again in 2010 when he attended my lecture about the development of astrometry and I was happy when he, my admired teacher, wrote to me: *"… unbelievable what you have achieved."*

Naur encouraged me in 1958 to go abroad. I objected that I had to learn more before going, but he said that I should go now exactly in order to learn. In his home in Brorfelde, Naur taught me to like operas sitting besides of me with the score pointing at the notes while we listened to his LPs (long playing disk) which were novel at the time and which he let me copy on tape.

Since Naur had meant so much for me I phoned him when I wrote this section and we had a very pleasant half-hour talk. He told for instance that he had completely forgotten, actually repressed the three years in Brorfelde because they had been so unpleasant and scientifically unproductive for him – all the while Brorfelde had meant so much for me. He has later read the present draft and agrees.

**Friends in Hamburg**

Mrs Anne Heinke was the secretary of Heckmann and she had immediately adopted me in her family on my arrival in October 1958. During the war she had lived in a city east of Dresden and fled with her daughter and two sons towards Dresden, away from the Russians. When they arrived at the city they were not let in but had to stay outside during the night and that saved their lives. In the night they saw the city in flames. It was the night in February 1945 when the allied air forces bombed Dresden and ten thousands died. I had never heard of that before Mrs Heinke told me. On Wikipedia I now read: *"In four raids between 13 and 15 February 1945, 722 heavy bombers of the British Royal Air Force (RAF) and 527 of the United States Army Air Forces (USAAF) dropped more than 3,900 tons of high-* *explosive bombs and incendiary devices on the city."*

In Heinke's home I soon met Herrn Nitschke who had just been released from the DDR political prison in Bautzen. The small man told with great and grim humour about his experiences. Entering Bautzen, he was surrounded by men who wanted to know how much he had got. At his answer "two years" they laughed and said, "Oh, then you have done nothing." They often had got five or ten years. His offence was that at a visit in West Berlin he had told that you have to be recorded (aufgeschrieben) if you want to buy butter in East Berlin. Someone must have heard him saying and had told the police. As soon as he was out after the two years he went to a shop wanting to buy butter. "Sind Sie aufgeschrieben?" they asked. So, nothing had changed and to tell that in the West was an offence against the *Volksrepublik*. Erik from nice and peaceful Denmark listened and we had a great time at Heinkes.

Heinke had been a soldier at the front and told that even two months before the end of the war he and the other soldiers heard a radio talk by Goebbels which convinced them that Germany would win.

Herr Heinke worked in a big tobacco machine factory in Bergedorf, Hauni, and he had obtained the task to photograph tobacco machines in Italy. He was going there in May 1960 with a Volkswagen transport wagon containing all his equipment, but still with room for one more person at his side on the front seat. I grasped the offer to be that person – and Heckmann said: *"don't ask me for permission, you are free!"*. Heinke had carefully chosen the route to pass all the cultural wonders on the way through Germany and Italy: e.g. Maria Laach, Verona, Bologna, Florens, San Gimignano, Siena, Viterbo, Rome, Orvieto, Spoleto, Assisi, Ravenna, Venice. I had never even heard of the wonderful churches in Florens and Ravenna, and forty years later I went again to show them to my wife, just the two of us, after the children had grown up.

With Heinke I had the time free to see all the eye-smashing cathedrals etc. while he was busy photographing tobacco machines. From him I learnt how to take good photos, coloured diapositives, and in the night we slept in the van or the grass after cooking our meal. I like to talk with people and one day in Rome I told the girl at my side in the bus that I was going to visit the Vatican, whereupon she offered to join me as a guide, and I accepted.

I stayed in the house of the family Gretel and Hans-Heinrich Voigt (Figure 21) for the first few years. Voigt came a year after me as Privatdozent to Hamburg from Göttingen and he stayed in Bergedorf until he returned to Göttingen as professor and director.

They had two children, Barbara and Christiane, the latter with a beautiful long hair which I was allowed to comb. The Voigts were very good to arrange for people to meet and we met in a literature group to read famous plays of Lorca, Brecht and Holberg with distributed roles, followed by discussion. In the circle were mostly Herczeg, professor Wellmann's, the Heinke's and soon also my wife Aase.

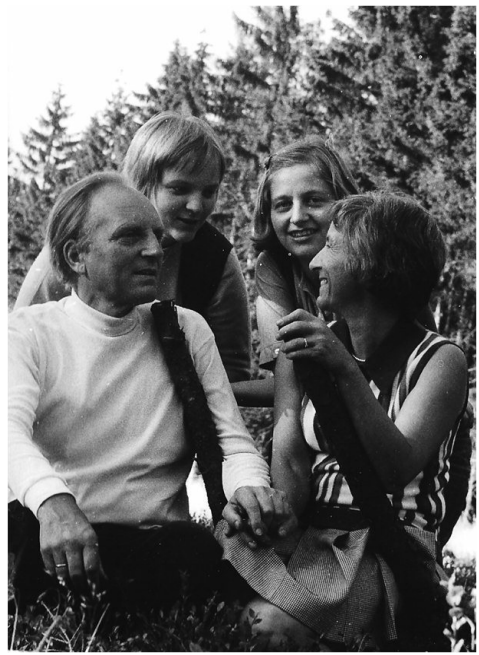

**Figure 21** Family Voigt in 1969. Hans-Heinrich and Gretel with Barbara and Christiane. - HV

The Hamburg University arranged for social opportunities for the foreign students. We could once a week come to a club where also German students frequented, the core was perhaps 20 or 30 persons. I have forgotten all names although this weekly event meant a great deal to me, we also had parties with dancing (Figure 22) and excursions to the beaches of the Baltic and the North Sea. One of the weekends we went by cars to a village near Celle where we were the guests of a farmer, an uncle of one of the students.

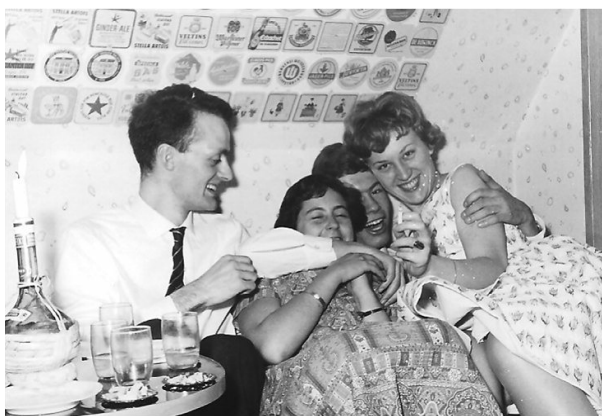

**Figure 22** Student party in Hamburg 1959. - EH

We were seated in a spacious garden room when the farmer welcomed us with a serious speech in which he regretted that we missed the great Führer, Adolf Hitler. So no doubt we were the guests of an old nazi, and he still was nazi – a very astounding situation for foreign students. We did not argue or discuss this issue ever, but we spent a wonderful weekend with the family. We danced, I loved to dance and girls liked to dance with me, we tried the Ratzeputz, a local schnaps, and I flew with the farmer from a local air field in a three-seated propeller of the amateur club of which he was a member.

The farmer and his family later came to the observatory on my invitation and I remember how impressed he was of what I showed him. Thus I acquainted with a nazi and saw that he was a good man, an important experience for me and unique during my stay in Germany. We visited them again another time.

I also visited the Esperanto club in Hamburg where they met in a small restaurant. We invited prof. M.G.J. Minnaert, director of the Utrecht Observatory and a friend of Esperanto with whom I always had a chat when we met at various occasions. He gave a lecture about Esperanto in the University attended by Heckmann, Haffner, Voigt and many others and followed by a meeting with the press. Minnaert firmly believed Esperanto would grow and play a great role in international communications. He compared the long process of adopting Esperanto as the international language with the process of adopting the meter system which took about a hundred years. – I learnt Esperanto in 1950. I am still fluent in the language and I have used it all over the world, Russia, Japan, America, China etc.

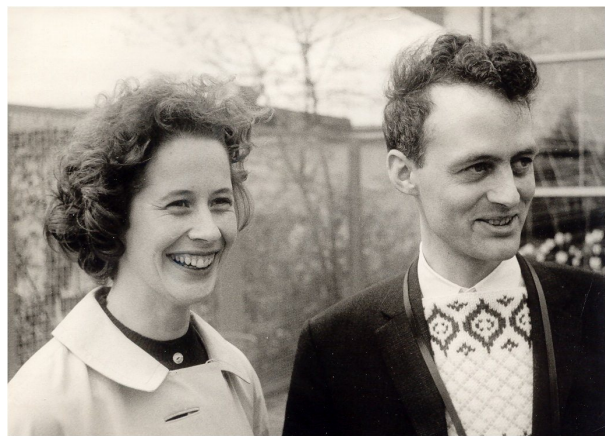

**Figure 23** Aase and Erik Høg 1963. - EH

**Dream girl and three children**

I had of course been looking for a girl in Hamburg although I wanted to marry a Danish girl. I did fall in love with a German beauty once, but after a short while she quit me. That luckily saved me for possible complications in such a combination, I later thought. In June 1961 I was invited to a party in Brorfelde arranged by my elder colleague Kjeld Gyldenkerne (1919-1999). I decided to drive back to Denmark just for that occasion. How good I did, for there I met my dream girl, Aase, then a school teacher in the nearby town Holbæk. So I found Aase, or she found me as you will see, and we married in December 1962 (Figure 23).

Aase came to Bergedorf in April 1963 and we moved into a small flat under the roof in the main building. This was offered by Heckmann although he had recently decided that it should not be used again for living, after the previous person had moved out. The flat had been installed in 1960 as guest room for the observatory. But he gave us the flat and my father and mother came from Denmark for a week to paint and prepare the first flat for their first married child. My father was a house painter and Aase learnt the metier partly from him and has practiced it ever since when we moved to a new place.

When we had moved into the flat in the observatory, Mrs Heckmann came to see us bringing a nice wooden bowl with delicious fruit. We were all very happy and grateful. When we had finished the fruit my parents thought we should bring back the wooden bowl, modest people as they were. Aase and I are also modest, but we kept the bowl and it decorated our living room with other fruits until a piece broke off a few years ago.

I saw both Heckmann's the last time in a rentiers' home in Göttingen in February 1979 and had the opportunity to tell him about the success of the Perth expedition and of the prospects for a first astrometric satellite, Hipparcos. They were both very glad to see me and Heckmann obviously felt how eager I was to tell him that I was grateful for the years in Bergedorf.

The occasion for my visit to Göttingen was a terrible snow storm when I left Copenhagen by train on the way to Carl Zeiss in Oberkochen. I was invited to discuss a new meridian circle they should build for the Tokyo Observatory. The storm brought so much snow that we were forced to stay overnight in Flensburg and were delayed by 24 hours on arrival in Hamburg. The S-Bahn could not go to Bergedorf at all where I had intended to speak with Kohoutek. I phoned family Voigt in Göttingen, and Gretel and Hans-Heinrich were glad to see me in the weekend after many years, and in the following week I went to Zeiss.

This was one of the heaviest snow falls ever in northern Germany. Heavy snow and a strong snow from the east produced large snowdrifts. For two to four days roads had not been useable, trains could not pass etc.

In 1964 our daughter, Birgitte, was born in Bethesda, a private clinic, where also our sons saw first light, Rasmus in 1966 and Torkil in 1970 (Figure 24). We found a new bigger flat in Lohbrügge some five kilometers from the observatory in an area which had been all farming when I first saw it in 1958. It soon developed into a nice small city with 20,000 inhabitants, and we built our house there in 1969. We lived close to the green area and there were many families and children to play with, but we have only few connections today since we moved to Denmark in 1973.

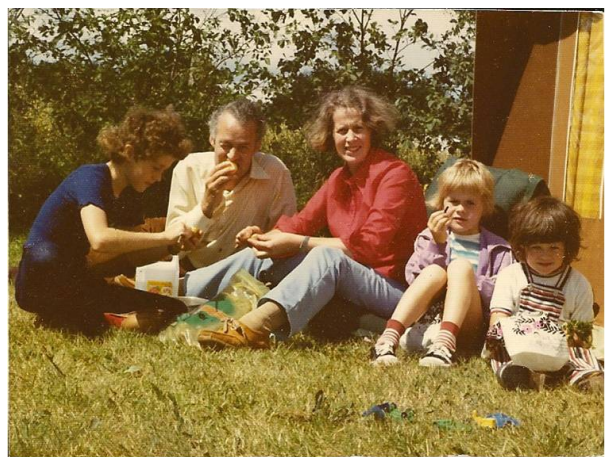

**Figure 24** Family Høg at their tent in 1974: Birgitte, Erik, Aase, Rasmus, and Torkil. - EH

When we had been married for over ten years, Aase had a "small" surprise for me one evening. She said that I had been subject to a conspiracy of two women, Inger Gyldenkerne and herself. She had asked Inger if they could invite me to the party in June 1961. She had seen me at the distance at my two visits to the drama club, but had never spoken with me. That may be called "attraction at a distance", and I had never seen her before she stood in front of me smiling when I came off the dancing floor. Nothing was more natural than to begin our first dance. When we came back in 1973 the observatory house offered to us happened to be the one where we had our first dance.

Soon however we bought a house, on Aase's initiative, in the nearby town Holbæk where we stayed for the next 21 years. Our children had moved out long ago and the house had become too big for us two elder persons. The house had then increased in value so that we could afford to buy a house near Copenhagen and we have since 1996 enjoyed to be close to all the Danish Capital offers, the Royal Theatre, exhibitions etc.

# 11. Forty years after

**Perth expedition A.D. 2010 and 2012**

How does the Perth expedition appear on the homepage of the Hamburger Sternwarte? That was easily found in February 2010 and a page under Research speaks in German of the *"development of the first photoelectric meridian circle telescope in the 1960s"* and the Perth70 Catalogue. The significance of the photoelectric method with slits and photon counting for the later Hipparcos satellite is not mentioned, nor is the importance of my experience with the development for my proposal in 1975 of a realistic design of Hipparcos. But without Otto Heckmann's immediate support of my ideas there would have come no Hipparcos. Quite naturally, this is not mentioned on the Hamburg website since even I have only begun to speak and write about these historical aspects in about 2008.

About the meridian circle five pages are found elsewhere on the website. A statement here seems a bit short: It says that *"1967 the meridian circle was dismounted, modernised and sent to Australia etc.",* as if the modernisation, which took the preceding seven years, was done during the packing.

Shortly after follows a section saying essentially this: The instrument was transferred to *Deutsches Museum* in Munich in 1989 where it should become part of the permanent exhibition on astronomy. When the exhibition was set up in 1992 the instrument was not included but placed in the depot where it still rests. It would have been very difficult to bring the instrument in a condition worthy for exhibition because in the course of time it had been modified by various "modern" devices and changed in its basic substance. Grundsubstanz is the German word used and the quotation marks around "modern" are placed as in the German text.

It appeared to me from the website as if *Deutsches Museum* was only interested in the original old version of a meridian circle for the exhibition, and not in the epoch making techniques for astrometry with photon counting developed in Hamburg although this technique and development became so crucial for the Hipparcos satellite, the satellite which started a new era of astrometry.

This was interesting to see because I think of my visit to *Deutsches Museum* in April 2003. I had a pleasant talk with the head of the division for astronomy Gerhard Hartl who thanked me for using my visit to Munich to come and see him. He said it was important for him to establish a connection with a participant of the expedition. I promised to send some material about the expedition, which of course I did shortly after my return to Copenhagen, the *Perth70 Catalogue* and various publications. This material may perhaps be found in the same boxes as the instrument in the depot of *Deutsches Museum*.

I can easily see the conflicting points of view facing a museum leader, so I take all this as an instructive example from real museum life. I should not blame anyone, but what you see in a science museum can be very far from the essentials of real science.

Later information supplied by J. Schramm in 2010: The observatory itself has thought about taking back the meridian circle into a science museum to be installed on the ground of the observatory, roughly around 1995 [or probably before 1992 when the instrument was brought back]. The museum should be installed inside the meridian circle building, but no money or institution could be found to form the museum.

This may however be realised if the observatory becomes part of a UNESCO world heritage project of historic observatories around the world. An application has been submitted, according to G. Wolfschmidt in 2013.

Two years later, in June 2012 the website shows apparently no change, but I understand from Professor Gudrun Wolfschmidt that the building has been renovated and that a new exhibition will be set up. The text and the exhibition will then – presumably - note the significance of the Hamburg meridian circle for the Hipparcos development. The present text of the website was mentioned above and reads in German: *"Es wäre auch sehr aufwendig gewesen, den Repsold-Meridiankreis in einen ausstellungswürdigen Zustand zu bringen, da er im Laufe der Zeit mit einigen "modernen" Geräte-Ergänzungen modifiziert bzw. in seiner Grundsubstanz verändert wurde."*

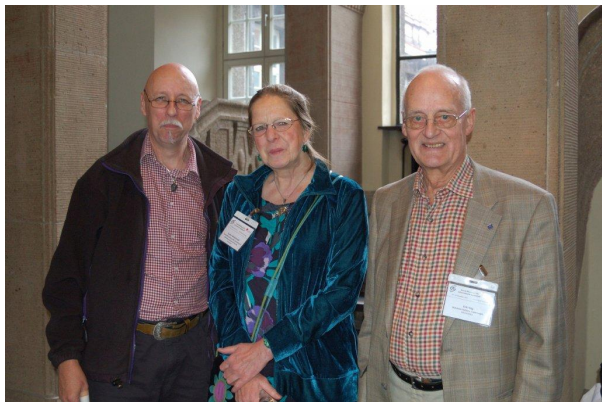

**Figure 25**  Craig Bowers, Gudrun Wolfschmidt and the author in Hamburg in 2012. - CB

In Perth, Mr. Craig Bowers (Figure 25) is working on a history of the Perth Observatory and I am happy that questions about the meridian circle instrument have been directed also to me. I had the pleasure in September 2012 to meet Craig on his visit to Hamburg and to see the nicely renovated meridian pavilion with him.

In March 2013 G. Wolfschmidt informs me: *...when the instrument was left over in Australia, no Hamburg institution was willing to spend money for bringing back the instrument; if our team for compiling the new astronomy exhibition in the Deutsches Museum (Teichmann, Hartl and I) would not have decided to save the instrument and to bring it to Munich (in addition we had just money for the exhibition), it wouldhave been completely distroyed.*

GW does not recognize or agree with the text quoted in German and she adds that the restoration itself was not discussed at that time. *There is in principle no problem in bringing the instrument into its latest important status which one would normally do.*

# 12. Astrophysicists and astrometry

**Miraculous approval of Hipparcos in 1980**

This report should close with the approval of the Hipparcos astrometry mission by ESA in 1980. Hipparcos met hard competition from the EXUV mission in the *Astronomy Working Group* (AWG) where the astrophysicists were in majority, and had Hipparcos failed in this advisory group there would have been no way to get into the decision process again, according to Professor Edward van den Heuvel. He was himself an X-ray astronomer with a stake in EXUV and had been asked by the chairman, his former boss Professor Cornelis de Jager, to present the Hipparcos proposal to the group. Quite unexpected by de Jager and not to his liking, to say the least, he then realized that van den Heuvel was very active in favour of Hipparcos.

Typically, astrophysicists know the value of astrometric data for the study of stars and the universe, but when it comes to the point of competition, their own project is more important. Details in this case are quoted by Høg (2011b: p.3) from the AWG meeting: At a meeting on 24 January 1980 the AWG considered the Astrometry and EXUV missions, concluding that both missions will give excellent scientific return. This is elaborated for the two missions. On astrometry for instance this: *"The Astrometry mission, HIPPARCOS, will give fundamental quantitative results to all branches of Astronomy. It emphasises typical European know how and will serve a community never before involved in space research"*; on the EXUV mission for instance this: *"The fact that the scientific objectives of this mission are being covered by two different missions proposed by other agencies (EUVE by NASA and ROBISAT by Germany) emphasises its timeliness."*

Even with such shaky argument five members still voted for EXUV! All five were brilliant scientists in their field of astrophysics. Of the thirteen members present, however, eight voted in favour of Hipparcos. Present at the meeting as members of AWG were thirteen persons: de Jager, Cezarsky, Delache, Drapatz, Fabian, Grewing, Jamar, Murray, Perola, Puget, Schilizzi, Spada, and Swanenburg while Rego was unable to attend, only Murray was an astrometrist. Van den Heuvel was no longer a member of AWG.

After the approval on 24 January, Hipparcos faced competition from the mission to Comet Halley, later called Giotto, which had been recommended by the Solar System Working Group. The final decision came in July when ESA had finally done two things it had never done before. The Scientific Programme Committee had approved two missions at the same time and it had decided that ESA should fund the Hipparcos payload out of the mandatory budget. Hitherto, payloads including instruments were always funded and built by the ESA member countries in national laboratories. In case the approval would have failed, Hipparcos or a similar scanning astrometry mission would never have been realized, neither in Europe nor anywhere else. The approval was "miraculous" for all these reasons as detailed in Høg (2011b).

Many years of work on astrometry had preceded the approval process as explained by Høg (2011a) and as briefly mentioned in the introduction to the present paper. The approval had never happened without the Hamburger Sternwarte, for this reason I have offered for exhibition there the model of Hipparcos received from ESA when I was consortium leader during the mission.

++++++++++++++++++++++++++++++++++

**Note about figures:**

The 25 Figures/photos here have different numbering than in the book edited by Gudrun Wolfschmidt. They are collected in a .ppt file (with **provisional** captions) and handouts are available:
https://dl.dropbox.com/u/49240691/AGAK12.ppt.pdf
https://dl.dropbox.com/u/49240691/AGAK25.ppt.pdf
++++++++++++++++++++++++++++++++++